\documentclass[amsmath,showpacs,nofootinbib,12pt]{revtex4-1}
\usepackage{graphicx}
\usepackage{dcolumn}
\usepackage{bm}
\usepackage{color} 
\usepackage{slashed}
\usepackage{float}
\usepackage{url}
\usepackage{caption}
\usepackage{hyperref}
\usepackage{cleveref}

\newcommand{\hs}{\hspace*{0.5cm}}

\newcommand{\be}{\begin{equation}}
\newcommand{\ee}{\end{equation}}
\newcommand{\bea}{\begin{eqnarray}}
\newcommand{\eea}{\end{eqnarray}}
\newcommand{\ben}{\begin{enumerate}}
\newcommand{\een}{\end{enumerate}}
\newcommand{\bde}{\begin{widetext}}
\newcommand{\ede}{\end{widetext}}

\newcommand{\crn}{\nonumber \\}

\newcommand{\al}{\alpha}
\newcommand{\la}{\lambda}

\newcommand{\ga}{\gamma}

\newcommand{\fr}{\frac}
\newcommand{\bc}{\begin{center}}
\newcommand{\ec}{\end{center}}
\newcommand{\Ga}{\Gamma}

\newcommand{\ka}{\kappa}
\newcommand{\La}{\Lambda}

\newcommand{\AdrHEPC}{Phenikaa Institute for Advanced Study (PIAS), Phenikaa University, Yen Nghia, Ha Dong, Hanoi 100000, Vietnam}
\newcommand{\AdrH}{Institute of Physics, Vietnam Academy of Science and Technology, 10 Dao Tan, Ba Dinh, Hanoi 100000, Vietnam}
\begin{document}
\title{Flavor-changing phenomenology in a $U(1)$ model} 
\author{N. T. Duy} 
\email{ntduy@iop.vast.vn}
\affiliation{\AdrH} 
\author{D. T. Huong}
\email{dthuong@iop.vast.vn (corresponding author)}
\affiliation{\AdrH} 
\author{Duong Van Loi}
\email{loi.duongvan@phenikaa-uni.edu.vn}
\affiliation{\AdrHEPC} 
\author{Phung Van Dong} 
\email{dong.phungvan@phenikaa-uni.edu.vn}
\affiliation{\AdrHEPC} 

\date{\today}

\begin{abstract}

We investigate a family-nonuniversal Abelian extension of hypercharge, which significantly alters the phenomenological features of the standard model. Anomaly cancellation requires that the third quark family transforms differently from the first two quark families. Additionally, it acquires that three right-handed neutrinos are presented. This model generates naturally small neutrino masses and a $W$-boson mass deviation appropriate to recent measurements. Additionally, the model introduces flavor-changing neutral currents (FCNCs) of quarks coupled to the new gauge boson $Z'$ and new Higgs fields. These FCNCs significantly modify the neutral-meson mixing amplitudes and rare meson decays, which are studied in detail.  We also address flavor changing processes in the charged lepton sector. 
\end{abstract}
\maketitle

\section{Introduction}

The evidence of neutrino oscillations ~\cite{McDonald:2016ixn,Kajita:2016cak} suggests that the Standard Model (SM) is incomplete. Moreover, the SM lacks a dark matter candidate, which makes up most of the mass of galaxies and galatic clusters, as already observed~\cite{Bertone:2004pz}.  

To address these shortcomings, one approach is to extend the SM with a $U(1)$ gauge symmetry. Popular approaches include $U(1)_{B-L}$, $U(1)_{T_{3R}}$, $U(1)_{L_i-L_j}$ for $i,j=e,\mu,\tau$, or a dark $U(1)_D$. Most of these extensions assume universality of quark and lepton families under the $U(1)$ charge, except for  $L_i-L_j$ model, which introduces non-universal lepton families.  In contrast,  the alternative $U(1)_X$ model \cite{VanDong:2022cin} proposes non-universal quark families. All lepton doublets and some quark doublets are assigned the same $X$ charge,  denoted by $x$, while the remaining quark doublets are assigned an opposite $X$ charge, i.e. $-x$. This model has been shown to explain the fermion family number as matched to the color number, because of anomaly cancellation. On the other hand, the theory can be embedded in a (flipped) trinification \cite{Huong:2016kpa,Dong:2017zxo,Singer:1980sw,Valle:1983dk,Pisano:1992bxx} or $E_6$ \cite{Glashow:1961tr} grand unification. Right-handed neutrinos appear as fundamental components, required by anomaly cancellation too. The breaking of this Abelian symmetry naturally induces small neutrino masses. 

The recent measurements of the $W$ boson mass have had significant implications for electroweak precision tests.
The SM prediction for the $W$ boson mass is based on well-established relationships between the electroweak coupling constants and the masses of $Z$ boson and top quark. Any deviation from these predictions could indicate to new physics beyond the SM. The extended Higgs sector as associated to a symmetry breaking beyond the SM leads to the mixing of neutral gauge bosons, $Z$ and $Z^\prime$. This reduces the $Z$ boson mass, thus it modifies the $W$-gauge boson mass. The combination of the electroweak precision tests on the $\rho$ parameter and the $W$ boson mass based on the CDF measurement provides constraints on the new physical scale \cite{VanDong:2022cin} and rules out the alternative $U(1)_X$ model according to $x=\pm \frac{1}{6}$. Notably,  the CDF experiment \cite{CDF:2022hxs} reported a surprisingly high precision measurement of the $W$ boson mass in 2022,  deviating significantly from the SM prediction. While  ATLAS \cite{ATLAS:2024erm} and LHCb \cite{LHCb:2021bjt} measurements have aligned with previous results, the most recent CMS measurement \cite{CMSMW2024}, 
\bea
m_W^{\text{CMS,2024}}=80.3602 \pm 0.0099 \ \text{GeV},
\eea
has a comparable precision to the SM prediction. The discrepancy between the CDF result and other measurements remains unresolved. However, the CMS measurements are  closer to the SM predictions, propping us to reassess earlier findings \cite{VanDong:2022cin}.

The third family of quarks transforms differently from the first two, leading to intriguing tree-level flavor-changing  neutral currents (FCNCs) mediated by the new $Z'$ gauge boson. Additionally, the model includes a new Higgs doublet, resulting in both up- and down-type quarks coupling to both Higgs doublets and inevitably giving rise to tree-level FCNCs mediated by the neutral Higgs field.  While the SM FCNCs vanish at tree level, they arise at the loop level through charged current contributions from $W$-boson, which are strictly suppressed by the GIM  mechanism. SM predictions for FCNC effects in meson physics, such as neutral-meson mixings and rare meson decays, $b \to s \gamma$, $t \to c \gamma$, are generally consistent with experimental observations \cite{Workman:2022ynf}. 
 
 Quark FCNCs impose strong constraints on new physics proposals \cite{Duy:2024lbd,Thu:2023xai,Duy:2022qhy,Dinh:2019jdg,Huong:2019vej}. Previous studies \cite{VanDong:2022cin} investigated the impact of the FCNCs associated with the new $Z'$ gauge boson on the meson mixing systems, providing qualitative lower bounds on the $Z'$ gauge boson mass. In this work, we carefully examine several  experimental observations related to the quark flavor-changing processes which are mediated by new gauge boson and scalar boson, considering both tree-level  and loop-level corrections mediated by new gauge bosons and scalar bosons. This comprehensive analysis allows us to derive  robust constraints on new physics scenarios.  Moreover, the model introduces three right-handed Majorana neutrinos, providing a solution for the observed smallness of neutrino masses. This also enriches the model's phenomenology with lepton flavor violation processes, making them a focal point of our research interest.
 
The rest of this work is organized as follows. In Sec. \ref{m1}, we give a review of the model in which the fermion, scalar, and gauge boson mass spectra are diagonalized, and the fermion couplings to scalars and gauge bosons are identified. In Sec. \ref{m2}, we revisit the electroweak precision fit complemented by the $W$ mass measurement. In Sec. \ref{m3}, we investigate the flavor-changing phenomenology. In Sec. \ref{m4}, we make a numerical analysis and discussion. Finally, we conclude this work in Sec. \ref{m5}.

\section{\label{m1} The model}
In the proposed model, the new electroweak group is $SU(2)_L \times U(1)_X \times U(1)_N$,  where  $U(1)_X \times U(1)_N$ breaks down to $U(1)_Y$ at high energy, with $N=Y-X$. The  X and 
  $N$ charges satisfy the condition: $3N_{q_L}+N_{l_L}=-(3X_{q_L}+X_{l_L}) \neq 0$, for each family.  
Since X and N can be unified with $SU(2)_L$ in a higher isospin group,  X and N  can be fixed as the neutral charges of the larger group \cite{VanDong:2022cin}. Specifically,  $X_{l_L}=x$ for all lepton doublets, $X_{q_L}=x$ for some quark doublets, and $X_{q_L}=-x$ for
 the remaining quark doublets. The  $[SU(2)_L]^2U(1)_X$ anomaly vanishes if the family number, $N_f =3(n-m)$, where $m$ and $n=N_f-m$ are the numbers of $q_L$ doublets with $x$ and $-x$, respectively. This implies that $N_f$ must be a multiple of the color number, leading to $N_f=3, n=2$, and $m=1$. All other anomalies, such as $[SU(3)_c]^2U(1)_X$, $[gravity]^2U(1)_X$,
 and others involving $X$ and $N$, also cancel out due to the matching number of families and colors and the presence of  the right-handed neutrinos.
  
  For symmetry breaking and mass generations, we introduce three scalar multiples:
 $H$ (identical to the SM Higgs doublet), $\Phi$ (coupling the two types of quarks to recover the CKM matrix), and $\chi$ (coupling to right-handed neutrinos for neutrino mass generation). The particle content of  the  model is listed in
 Table \ref{tab1}. 
  
 \begin{table}
 	\bc
 	\begin{tabular}{lcccc}
 		\hline\hline 
 		Field & $SU(3)_C$ & $SU(2)_L$ & $U(1)_X$ & $U(1)_N$\\
 		\hline 
 		$l_{aL}=\begin{pmatrix}
 			\nu_{aL}\\
 			e_{aL}\end{pmatrix}$ & 1 & 2 & $x$ & $-1/2-x$ \\
 		$\nu_{aR}$ & 1 & 1 & $x$ & $-x$\\
 		$e_{aR}$ & 1 & 1 & $x$ & $-1-x$\\
 		$q_{\al L}=\begin{pmatrix}
 			u_{\al L}\\
 			d_{\al L}\end{pmatrix}$ & 3 & 2 & $-x$ & $1/6+x$ \\
 		$u_{\al R}$ & 3 & 1 & $-x$ & $2/3+x$\\
 		$d_{\al R}$ & 3 & 1 & $-x$ & $-1/3+x$\\
 		$q_{3 L}=\begin{pmatrix}
 			u_{3 L}\\
 			d_{3 L}\end{pmatrix}$ & 3 & 2 & $x$ & $1/6-x$ \\
 		$u_{3 R}$ & 3 & 1 & $x$ & $2/3-x$\\ 
 		$d_{3R}$ & 3 & 1 & $x$ & $-1/3-x$\\
 		$H=\begin{pmatrix}
 			H^+_1\\
 			H^0_2
 		\end{pmatrix}$ & 1 & 2 & 0 & $1/2$\\
 		$\Phi=\begin{pmatrix}
 			\Phi^+_1\\
 			\Phi^0_2
 		\end{pmatrix}$ & 1 & 2 & $-2x$ & $1/2+2x$\\
 		$\chi$ & 1 & 1 & $-2x$ & $2x$\\
 		\hline\hline
 	\end{tabular}
 	\caption[]{\label{tab1} Field representation under the gauge symmetry.}
 	\ec
 \end{table}    
\subsection{Mass spectrum}  
\subsubsection{Scalar mass spectrum} 
 The scalar potential has a following form
 \bea
 V &=& \mu^2_1 H^\dagger H + \mu^2_2 \Phi^\dagger \Phi + \mu^2_3 \chi^\dagger \chi +\mu_4 [(\Phi^\dagger H) \chi+H.c.] \crn
 &&+ \la_1(H^\dagger H)^2+\la_2(\Phi^\dagger \Phi)^2+\la_3 (\chi^\dagger \chi)^2\crn
 && +\la_4 (H^\dagger H)(\chi^\dagger \chi) +\la_5 (\Phi^\dagger \Phi) (\chi^\dagger \chi)\crn
 &&+ \la_6 (H^\dagger H)(\Phi^\dagger \Phi) + \la_7(H^\dagger \Phi) (\Phi^\dagger H), \label{Scalar1} \eea   
where $\tilde{H}=i\sigma_2 H^*$, $\tilde{\Phi}=i \sigma_2 \Phi^*$, and $f^\nu$, $h$'s, and $\la$'s are dimensionless, while $\mu_{1,2,3,4}$ have a mass dimension. After symmetry breaking, the scalar fields have the  vacuum expectation values (VEVs) are given by 
\bea &&\langle \chi\rangle =\La/\sqrt{2},\space \hs  \langle H\rangle = (0,v_1/\sqrt{2}),\space \hs  \langle \Phi \rangle =(0,v_2/\sqrt{2})\eea 
This satisfies $v^2_1+v^2_2=(246\ \mathrm{GeV})^2$ for consistency with the SM. To understand its stability and vacuum structure, we need to determine the necessary conditions for this potential to be:
\bea &&\mu^2_{1,2,3}<0,\hs \la_{1,2,3}>0, \hs |\mu_{1,2}|\ll |\mu_3|\\
&& \la_4>-2\sqrt{\la_1\la_3},\hs \la_5>-2\sqrt{\la_2\la_3},\hs \la_6+\la_7\Theta(-\la_7)>-2\sqrt{\la_1\la_2}. \eea
We expand the scalar fields around their VEVs as follows:
\bea H &=& \begin{pmatrix}
	H^+_1\\
	\fr{1}{\sqrt2}(v_1+S_1+iA_1)
\end{pmatrix},\hs \Phi=\begin{pmatrix}
	\Phi^+_1\\
	\fr{1}{\sqrt2}(v_2+S_2+iA_2)
\end{pmatrix},\\
\chi &=& \fr{1}{\sqrt2}(\La+S_3+iA_3),\eea
Substituting these fields into the scalar potential, we obtain the following minimum potential conditions:
\bea (2\la_1v_1^2+\la_4\La^2+\la_6v_2^2+\la_7v_2^2+2\mu_1^2)v_1+\sqrt2\La\mu_4 v_2 &=& 0,\\
(2\la_2v_2^2+\la_5\La^2+\la_6v_1^2+\la_7v_1^2+2\mu_2^2)v_2+\sqrt2\La\mu_4 v_1 &=& 0,\\
(2\la_3\La^2+\la_4v_1^2+\la_5v_2^2+2\mu_3^2)\La+\sqrt2\mu_4 v_1 v_2 &=& 0.
\label{scalar2}\eea
Utilizing the conditions derived in Eqs.(\ref{scalar2}),  we obtain a predicted scalar mass spectrum for considered model. This spectrum comprises three massive neutral CP-even Higgs bosons, denoted by $\mathcal{H}, \mathcal{H}_1, \mathcal{H}_2$ .  In the limit, $\mu_4 \simeq \La \gg v_1, v_2$, they have the following masses:
\bea m^2_{\mathcal{H}}&\simeq& \fr{2[\la_1v_1^4+\la_2v_2^4+(\la_6+\la_7)v_1^2v_2^2]}{v_1^2+v_2^2}-\fr{(\la_4v_1^2+\la_5v_2^2)^2}{2\la_3(v_1^2+v_2^2)}\\
&&-\fr{\sqrt2 v_1v_2(\la_4v_1^2+\la_5v_2^2)}{\la_3(v_1^2+v_2^2)}\fr{\mu_4}{\La}-\fr{v_1^2v_2^2}{\la_3(v_1^2+v_2^2)}\fr{\mu_4^2}{\La^2},\\
m^2_{\mathcal{H}_1}&\simeq&-\frac{\mu_4\La(v_1^2+v_2^2)}{\sqrt2 v_1v_2},\hs m^2_{\mathcal{H}_2}\simeq2\la_3\La^2. \eea
There are also CP-odd and charged Higgs bosons denoted by $\mathcal{A}$ and $\mathcal{H}^\pm$. Their masses are also given by 
\bea m^2_\mathcal{A}&=&-\fr{\mu_4[\La^2(v_1^2+v_2^2)+v_1^2v_2^2]}{\sqrt2\La v_1v_2},\\ m^2_\mathcal{H^\pm}&=&-\fr{(\sqrt2\La\mu_4+\la_7v_1v_2)(v_1^2+v_2^2)}{2v_1v_2}.
\eea
The model predicts the existence of massless particles called Goldstone bosons, denoted by $G_Z, Z_{W^\pm}, G_{Z^\prime}$. These particles are absorbed by  $Z, W^\pm, Z^\prime$ gauge boson.

In term of physical states, the original Higgs doublets and singlet Higgs can be written as linear combinations of physical fields. These combinations can be expressed as:

\bea H &\simeq& \begin{pmatrix}
	c_\al G^+_W+s_\al \mathcal{H}^+\\
	\fr{1}{\sqrt2}(v_1+c_\al\mathcal{H}+s_\al\mathcal{H}_1+ic_\al G_Z+is_\al\mathcal{A})
\end{pmatrix},\\
\Phi&\simeq&\begin{pmatrix}
	s_\al G^+_W-c_\al \mathcal{H}^+\\
	\fr{1}{\sqrt2}(v_2+s_\al\mathcal{H}-c_\al\mathcal{H}_1+is_\al G_Z-ic_\al\mathcal{A})
\end{pmatrix},\\
\chi &\simeq& \fr{1}{\sqrt2}(\La+\mathcal{H}_2+iG_{Z'}).\eea
 Here, we define $\tan \al \equiv t_\al =\frac{v_2}{v_1}$.  
 \subsubsection{Fermion masses}
 The fermion masses are analyzed in more  detail in \cite{VanDong:2022cin}. For the reader's convenience, we summarize the key findings.  The Yukawa Lagrangian of the model can be expressed as \cite{VanDong:2022cin} 
\bea \mathcal{L}_{\mathrm{Yuk}} &=& h^e_{ab}\bar{l}_{aL} H e_{bR}+h^\nu_{ab} \bar{l}_{aL} \tilde{H} \nu_{bR}+\fr 1 2 f^\nu_{ab} \bar{\nu}^c_{a R} \nu_{bR}\chi\crn
&&+ h^d_{\al \beta} \bar{q}_{\al L} H d_{\beta R}+ h^u_{\al \beta} \bar{q}_{\al L} \tilde{H} u_{\beta R} +h^d_{33}\bar{q}_{3L} H d_{3R}+ h^u_{33} \bar{q}_{3L} \tilde{H} u_{3R} 
\crn
&&+ h'^d_{\al 3} \bar{q}_{\al L} \Phi d_{3 R}+ h'^u_{3\beta } \bar{q}_{3L} \tilde{\Phi} u_{\beta R} +H.c.
\label{Yuka1}\eea
After the gauge symmetry breaking, fermions obtain their masses, as discussed in \cite{VanDong:2022cin}. The d-quarks, $(d_1,d_2,d_3)$, mix together according to the  mass matrix:
\bea
M_{d}= -\frac{v_1}{\sqrt{2}} \begin{pmatrix}h_{11}^d & h_{12}^d & t_\al h_{13}^{\prime d} \\
	h_{21}^d & h_{22}^d & t_\al h_{23}^{\prime d} \\ 0 & 0 & h^d_{33},
\end{pmatrix}, 
\eea
and in the the basis $(u_1,u_2,u_3)$, the mass matrix has the form
\bea
M_{u}= -\frac{v_1}{\sqrt{2}} \begin{pmatrix}h_{11}^u & h_{12}^u & 0 \\
	h_{21}^u & h_{22}^u & 0 \\ t_\al h_{31}^{\prime d} & t_\al h_{32}^{\prime d} & h^u_{33},
\end{pmatrix}. 
\eea
The quark matrices, $M_u, M_d$, can be diagonalized by a pair of bi-unitary matrices:
    \bea
    V_{uL}^\dag M_u V_{uR} = \text{Diag}\begin{pmatrix} m_u & m_c, m_t\end{pmatrix}; \hs 
    V_{dL}^\dag M_d V_{dR} = \text{Diag}\begin{pmatrix} m_d & m_s, m_b\end{pmatrix}.
    \eea
 Charged leptons acquire mass through their coupling,$h^e$, given by: \be [m_e]_{ab}=-h^e_{ab}\fr{v_1}{\sqrt{2}} \label{mass_lep}.\ee 
 The smallness of  active neutrinos is explained by the canonical seesaw mechanism, involving  right-handed neutrinos with large Majorana masses, $[m_M]_{ab}=-f^\nu_{ab}\fr{\La}{\sqrt{2}}$, and  Dirac mass via $h^\nu$ coupling, $[m_D]_{ab}=-h^\nu_{ab} \fr{v_1}{\sqrt{2}}$. Active neutrinos thus acqurire a small effective mass given by:
\be [m_{\nu_l}]_{ab}= -[m_D m^{-1}_M m^T_D]_{ab} \sim v^2_1/\La,\ee 
while heavy neutrinos have a mass:
\be
m_{\nu_h} \simeq M.
\ee
The physical neutrino states $\nu_l, \nu_h$  are related to the flavor states via a rotation matrix:
\bea
\begin{pmatrix} \nu_L \\ \nu_R^c \end{pmatrix} = V_\nu \begin{pmatrix} \nu_l \\ \nu_h \end{pmatrix}\equiv \begin{pmatrix} V_{Ll} & V_{Lh} \\ V_{Rl} & V_{Rh}  \end{pmatrix}\begin{pmatrix} \nu_l \\ \nu_h \end{pmatrix}.
\eea
\subsubsection{Gauge boson masses}
Let us list the gauge boson mass spectrum, as considered in \cite{VanDong:2022cin}. Gauge boson masses arise from the scalar kinetic terms $\sum_s \left( D^\mu S\right)^\dag \left(D_\mu S \right)$ after spontaneous symmetry breaking.  Similar to the SM, the model predicts  the physical charged gauge boson, $W^\pm _\mu =\frac{A_{1\mu}\mp i A_{2 \mu}}{\sqrt{2}},$ with mass $m_W^2 =\frac{g^2}{4}\left( v_1^2+v_2^2\right)$.  Three neutral gauge bosons, $A_{3 \mu}, B_\mu, C_\mu$, mix.  After a suitable rotation,, we obtain the photon $A_\mu$ and two physical sates, denoted by $Z_1, Z_2$,  given by:
\bea
Z_1= \cos \varphi Z -\sin \varphi Z^\prime, \hspace{1cm} Z_2= \sin \varphi Z +\cos \varphi Z^\prime,
\eea
where
\bea
\tan 2 \varphi  \simeq \frac{\sin 2 \theta \left(\sin^2 \theta v_1^2+(\sin^2 \theta+4x)v_2^2 \right)}{16 s_W x^2 \La^2}
\eea  
with
\bea
A_\mu &= & \sin \theta_W A_{3 \mu} +\cos \theta_W\left(\sin \theta_W B_\mu + \cos \theta  C_\mu\right), \crn \nonumber 
Z_\mu &=& \cos \theta_W A_{3\mu}-\sin \theta_W\left(\sin \theta B_\mu + \cos\theta C_\mu\right), \crn \nonumber 
Z^\prime_\mu &= & \cos\theta B_\mu -\sin \theta C_\mu.
\eea
Here, $\tan \theta_W=\frac{g_Y}{g}=\frac{g_Xg_N }{g \sqrt{g_X^2+g_N^2}}$, $\tan \theta =\frac{g_N}{g_X}.$  
\subsection{Scalar and vector currents}
\subsubsection{(Pseudo) Scalar currents}
The Yukawa interactions can be split into the fermion masses and interaction terms. These interaction terms have a form
\bea
\mathcal{L}_{\text{int}}^s= \mathcal{L}_{\text{int}}^{s-\text{NC}}+\mathcal{L}_{\text{int}}^{s-\text{CC}},
\eea
where $\mathcal{L}_{\text{int}}^{s-\text{NC}}$  is neutral scalar currents for both lepton and quark. In the limit $\mu_4 \simeq \La \gg v_1,v_2$,  it can be written as follows:
\bea
\mathcal{L}_{\text{int}}^{s-\text{NC}}=\mathcal{L}_{l-s}^{\text{NC}}+
\mathcal{L}_{q-s}^{\text{NC}},
\eea
where 
\bea
\mathcal{L}_{l-s}^{\text{NC}}=-\frac{g}{2m_W}\left\{\bar{e}_{aL} m_{l_a} e_{aR} \left\{\mathcal{H}+t_\al \left( \mathcal{H}_1+i \mathcal{A}\right) \right\}-\bar{\nu}_{aL} m_{\nu_a}^D \nu_{aR} \left\{\mathcal{H}+t_\al \left( \mathcal{H}_1+i \mathcal{A}\right) \right\} \right\}+H.c., \nonumber \\
\eea 
and 
\bea
\mathcal{L}_{q-s}^{\text{NC}}=&&-\frac{g}{2 m_W}\left\{\sum_i \bar{q^\prime}_{iL}m_{q^\prime_i}q^\prime_{iR}\mathcal{H}+t_\al\sum_i \bar{q^\prime}_{iL}m_{q^\prime_i}q^\prime_{iR}\mathcal{H}_1+\sum_{i,j} \bar{q^\prime}_{iL}\left(\Ga^q_{\mathcal{H}_1}\right)_{ij}q^\prime_{jR}\mathcal{H}_1 \right\} \crn  && -i\frac{g}{2 m_W}\left\{\pm t_\al \sum_i \bar{q^\prime}_{iL}m_{q^\prime_i}q^\prime_{iR} \mathcal{A}\pm\sum_{i,j} \bar{q^\prime}_{iL}\left(\Ga^q_{\mathcal{A}}\right)_{ij}q^\prime_{jR}\mathcal{A}\right\}+ H.c.
\label{fcnc_higgs1}\eea
with
\bea
&& \left(\Ga_{\mathcal{H}_1}^d \right)_{ij}=  \frac{-2}{s_{2\al}}\sum_{\beta=1}^2\sum_{k=1}^3\left(V_L^d \right)^\dag_{i\beta}\left( V_L^d\right)_{\beta k}m_{d_k}\left( V_R^d\right)^\dag_{k3}\left( V_R^d\right)_{3j}, \crn 
&& \left(\Ga_{\mathcal{H}_1}^u \right)_{ij}=  
-\fr{2}{s_{2\al}} \sum_{\beta=1}^2\sum_{k=1}^3\left(V_L^u \right)^\dag_{i3}\left( V_L^u\right)_{3 k}m_{u_k}\left( V_R^u\right)^\dag_{k\beta}\left( V_R^u\right)_{\beta j}.\nonumber
\eea
taking $-$ for u-quark and $+$ for d-quark, and $ \Ga_{\mathcal{A}}^q= \Ga_{\mathcal{H}_1}^q=\Ga^q$. 
In equation (\ref{fcnc_higgs1}), the terms containing interaction constants $ \Ga_{\mathcal{A}}^q, \Ga_{\mathcal{H}_1}^q$, and $\Ga^q$ 
describe FCNC interactions. In contrast, the remaining terms in the equation represent interactions that conserve flavor, 
similar to the coupling observed the SM. Note that the charged leptons couples only the SM Higgs doublet, it
does not encounter any tree-FCNC for them. 
The scalar charged currents can be written as:
\bea
\mathcal{L}_{\text{int}}^{s-\text{CC}}=\mathcal{L}_{l}^{s-\text{CC}}+\mathcal{L}_{q
}^{s-\text{CC}}
\eea
with
\bea
\mathcal{L}_{q
}^{s-\text{CC}}&& =-\fr{g}{\sqrt{2}m_W}\left\{\bar{d}^{\prime}_{iL}\mathcal{X}_{ij}u^\prime_{jR}+\bar{d}^{\prime}_{iR}\ \mathcal{Y}_{ij}  u^\prime_{jL}\right\} \mathcal{H}^{-} +H.c., \label{bs1}
\eea
where 
\bea 
\mathcal{X}_{ij}& =& -t_{\al} (V_{d_L}^{\dagger})_{ij}m_{u_j} +\sum_{k=1}^3\sum_{\beta=1}^2\left[\fr{2}{s_{2\al}}(V_{d_L}^{\dagger})_{i3}(V_{u_L})_{3 k}m_{u_k}(V_{u_R}^{\dagger})_{k\beta}(V_{u_R})_{\beta j}\right], \crn   
\mathcal{Y}_{ij}& =& t_{\al} m_{d_i}(V_{d_L}^{\dagger})_{ij}- \sum_{k=1}^3\sum_{\beta=1}^2\left[\fr{2}{s_{2\al}}(V_{d_R}^{\dagger})_{i3}(V_{u_R})_{3 k}m_{d_k}(V_{d_L}^{\dagger})_{k\beta}(V_{u_L})_{\beta j}\right].\eea
The charged scalar currents associated to lepton have a form:

\begin{align}
\mathcal{L}_{l
}^{s-\text{CC}}=	-\frac{g}{\sqrt{2}m_W}t_\alpha \left\{\bar{\nu}_l \left(- m_{\nu_l}V^\dagger_{Rll}P_L+V_{Ll}^\dag m_l P_R\right) l \right. \\
	\quad \left.+\bar{\nu}_h \left(- m_{\nu_h}V^\dagger_{Rh}P_L+ V^\dagger_{Lh}m_l P_R \right) l
\right\}\mathcal{H}^++ H.c.
\label{bs2}\end{align}
Charged scalar currents introduce new contributions to flavor-changing processes in both the lepton and quark sectors at the loop level. We will delve deeper into these contributions in the next section.
\subsubsection{(Axial) Vector Currents}

The $W^\pm$ interact with fermions similarly as the SM 
\bea
\mathcal{L}_W = -\frac{g}{\sqrt{2}} \left\{ \left( \bar{\nu}_L V^\dag_{Ll}+\bar{\nu}_h V^\dag_{Lh}\right) \gamma^\mu P_L l  \right\} W_\mu^+ + H.c.
\eea
The interactions of neutral gauge bosons $Z_{1,2}$ with fermions have been given in \cite{VanDong:2022cin}. Quark families transform differently under $X$ and $N$, but transform identically under $T_3$ and hypercharge $Y$. Ignoring the $Z-Z^\prime$ mixing, the model provides FCNCs coupled to the $Z^\prime$ gauge boson. These FCNCs can be expressed mathematically as follows:
\be \mathcal{L}_{\mathrm{q-Z^\prime}}=L_{ij} g \bar{q^\prime}_{iL}\ga^\mu q^\prime_{jL} Z'_\mu + (L\rightarrow R),\ee where we have defined,
\be L_{ij}=-2x\fr{t_W}{s_\theta c_\theta } (V^*_{qL})_{3i}(V_{qL})_{3j}. \label{fcnczp}\ee 

\section{\label{m2} Electroweak fit with the CMS $W$ boson mass measurement}
\begin{figure}[H]
	\centering
\begin{tabular}{cc}
		\includegraphics[width=7.5cm]{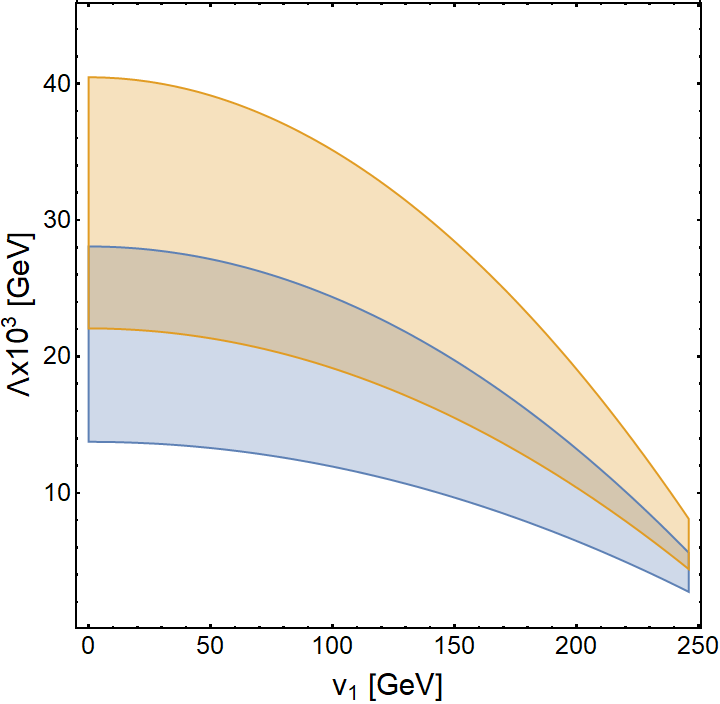}&
		\includegraphics[width=7.5cm]{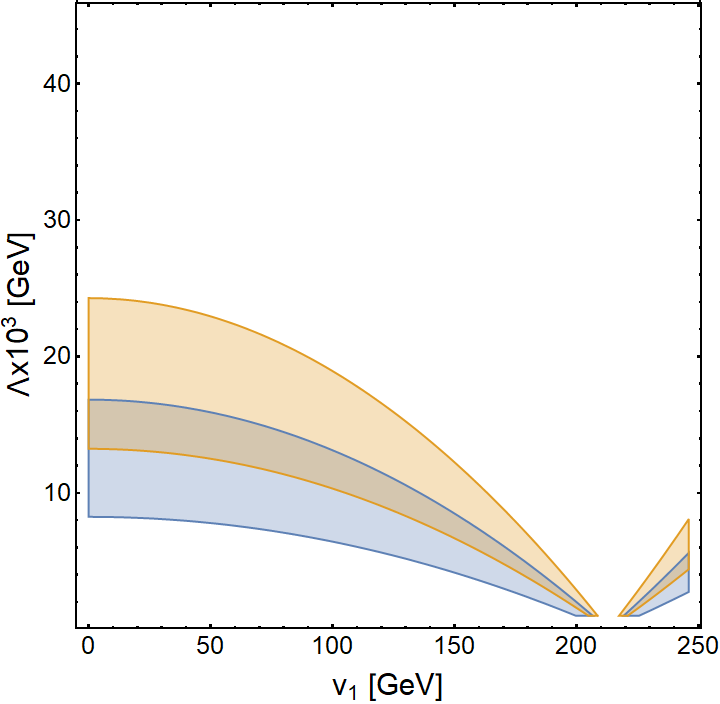}\\ 
			\includegraphics[width=7.5cm]{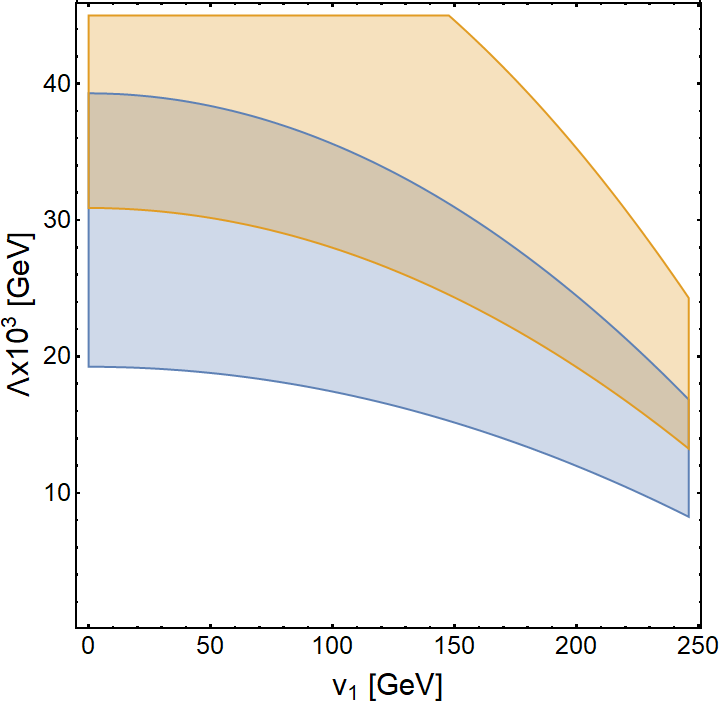}&
		\includegraphics[width=7.5cm]{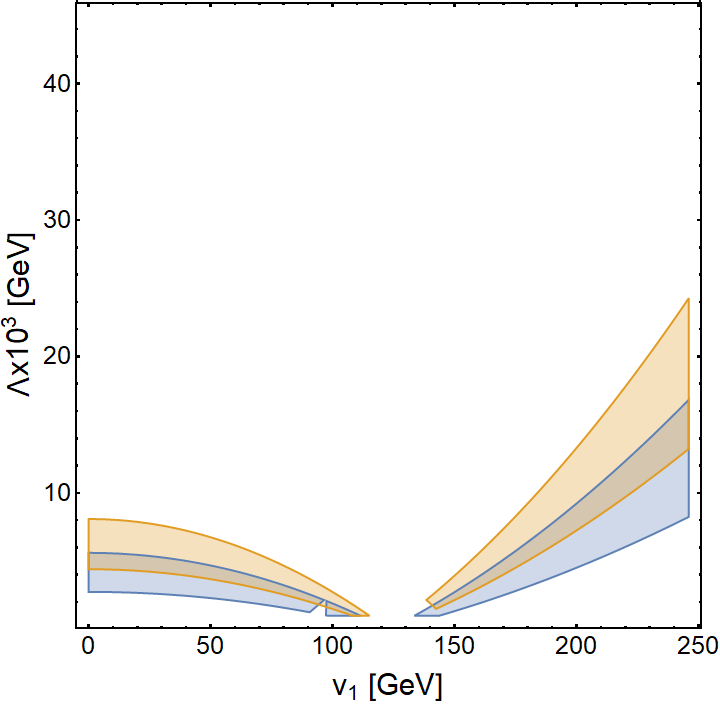}
	\end{tabular}
	\caption{ \label{MWCMS}
		The blue and yellow areas represent the region allowed by the deviation between CMS measurement and SM result of the W-mass $\Delta m_W^2$ and electroweak precision test on $\rho$ parameter  \cite{Workman:2022ynf}, respectively, in the $v_1-\La$ plane. The top and bottom panels correspond to the model with $x=\pm \frac{1}{2}$ and  $x=\pm \frac{1}{6}$.}
\end{figure}

The measurement of the $W$ boson mass has significant implications for electroweak precision tests.
In the previous studies  \cite{VanDong:2022cin}, the authors used CDF measurement \cite {CDF:2022hxs} for  the $W$ boson mass in combination with electroweak precision tests on $\rho$ parameter. However, given the recent discrepancies between the CDF data and LHC measurements  reported by CMS and ATLAS, the validity of the model under consideration is subject to scrutiny. To address this, we will re-examine the new physics contributions to gauge boson masses, focusing on the LHC measurement.  The mixing between $Z$ and $Z'$ can potentially reduce the observed $Z_1$ mass compared to the SM $Z$ boson mass. It can also give rise to a positive contribution to $\rho-$ parameter \cite{VanDong:2022cin}:  
\bea
\rho -1 = \frac{m_W^2}{c_W^2 m_{Z_1}^2}-1 \simeq  \tan^2 2 \varphi \simeq \al T.
\eea 
It leads to a dominant enhancement of the $W$ mass:
\bea
\Delta m^2_W\simeq \frac{\cos^4 \theta_W m_Z^2}{\cos 2\theta_W} \al T
\eea 
The Fig. \ref{MWCMS} shows the regions allowed by both the deviation between CMS measurement with SM prediction of the $W$ boson mass $\Delta m_W^2=m_W^2|_{\text{CMS}}-m_W^2|_{\text{SM}}=1.1571\pm0.6269$ $\text{GeV}^2$ and electroweak precision test on $\rho$ parameter $\rho=1.00031\pm0.00019$ \cite{Workman:2022ynf}. The figures are plotted in the $v_1-\La$ plane for different values $x=\pm \frac{1}{2}, \pm \frac{1}{6}$, whereas other parameters are chosen as given in \cite{VanDong:2022cin}, i.e $t_{\theta}=1$, 
and $s_W^2\simeq 0.231$. The parameter space, particularly dependent on the $x$ charge, has significantly shifted compared to previous results  \cite{VanDong:2022cin}. All four models, $x=\pm \frac{1}{2}, \pm \frac{1}{6}$, allow for a region of parameter space consistent with the CMS measurements of the $W$ boson mass and global fits of the $\rho$ parameter. For each value of $v_1$, we can determine the permissible  range of the new physical scale $\La$. For example, we find $\La=4.3-28.1$ TeV for $x=1/2$ and $\La=13.2-39.3$ TeV  and $x=1/6$. However, the model with $x=-\frac{1}{2}$ and $x=-\fr{1}{6}$ impose specific constraints on $v_1$, namely  $\La=1-16.7$ TeV, $v_1\leq 204$ GeV or $\La=1-5.4$ TeV, $v_1\geq 220$ GeV for $x=-1/2$; $\La=1-5.7$ TeV, $v_1\leq 108$ GeV or $\La=1.5-16.7$ TeV, $v_1\geq 141$ GeV for $x=-1/6$. 
 
\section{\label{m3} Flavor phenomenology}
\subsection{Quark flavor phenomenology}
\subsubsection{Neutral meson mixing}
Previous studies \cite{VanDong:2022cin} have explored the specific contribution of vector FCNCs to the amplitude of neutral meson mixing  while neglecting the contribution of scalar currents. In this work, we aim to analyze the combined impact of both vector and scalar interactions on the mass difference of the mesons. From the scalar FCNCs given in ( \ref{fcnc_higgs1}) and vector FCNCs in (\ref{fcnczp}), we can construct the effective interaction involving four quarks by integrating out the heavy fields $\mathcal{H}_1, \mathcal{H}_2, \mathcal{A}$, and $Z^\prime$. This process yields the effective interactions,  
\bea \mathcal{H}^{\mathrm{eff}}_{\mathrm{FCNC}} &&= -\frac{g^2}{4m_W^2}\left\{\left(\Ga_{ij}^q\right)^2\left(\frac{1}{m_{\mathcal{H}_1}^2}-\frac{1}{m_{\mathcal{A}}^2} \right)\left(\bar{q}_{iL}q_{jR}\right)^2+ \left(\Ga^{q*}_{ji} \right)^2\left(\frac{1}{m_{\mathcal{H}_1}^2}-\frac{1}{m_{\mathcal{A}}^2} \right)\left(\bar{q}_{iR}q_{jL} \right)^2\right\} \crn \nonumber &&- \frac{g^2}{4m_W^2}\left\{\Ga_{ji}^{q*}\Ga_{ij}^q\left(\frac{1}{m_{\mathcal{H}_1}^2}+\frac{1}{m_{\mathcal{A}}^2}\right)
\left\{(\bar{q}_{iL}q_{jR})(\bar{q}_{iR}q_{jL}) +(\bar{q}_{iR}q_{jL})(\bar{q}_{iL}q_{jR})\right\}\right\}  \crn \nonumber &&+ \fr{g^2}{m^2_{Z_2}} \left\{ L^2_{ij}(\bar{q}_{iL}\ga^\mu q_{jL})^2+ 2L_{ij}R_{ij}(\bar{q}_{iL}\ga^\mu q_{jL})(\bar{q}_{iR}\ga^\mu q_{jR})+R^2_{ij}(\bar{q}_{iR}\ga^\mu q_{jR})^2\right\}.\label{ptdt11}\eea 
The effective interactions presented in Eq. (\ref{ptdt11}) contribute to the amplitude of neutral meson mixing. The calculations yield the following results:
\bea
\Delta m_K^{\text{NP}} 
&&\simeq \fr{2}{3} \mathrm{Re}\left\{L^2_{12}-\left[\fr 3 2+ \left(\fr{m_K}{m_d+m_s}\right)^2\right]L_{12}R_{12}+R^2_{12}\right\} \frac{g^2}{m^2_{Z_2}}m_K f^2_K \nonumber \crn && +\frac{5}{48}\mathrm{Re} \left\{\left[ \left(\Ga_{12}^q\right)^2+ \left(\Ga_{21}^{q*}\right)^2\right]\left(\frac{m_K}{m_s+m_d} \right)^2\right\} \left(  \fr{g^2}{m^2_{\mathcal{H}_1}}-
\frac{g^2}{m^2_\mathcal{A}} \right) m_K f_K^2 \nonumber \crn && - \frac{1}{4}\mathrm{Re}\left\{ \Ga_{21}^{q*}\Ga_{12}^q  \left(\frac{1}{6}+\frac{m_K^2}{\left(m_s+m_d\right)^2} \right)\right\}\left(\frac{g^2}{m^2_{\mathcal{H}_1}} +\frac{g^2}{m_{\mathcal{A}}^2}\right) m_K f_K^2,\label{ptdt16}
\eea 
\bea \Delta m_{B_d}^{\text{NP}} &&\simeq \fr{2}{3} \mathrm{Re}\left\{ L^2_{13}-\left[\fr 3 2+ \left(\fr{m_{B_d}}{m_d+m_b}\right)^2\right]L_{13}R_{13}+R^2_{13}\right\}\fr{g^2}{m^2_{Z_2}}m_{B_d} f^2_{B_d}\nonumber \crn
&& +\frac{5}{48}\mathrm{Re} \left\{\left[ \left(\Ga_{13}^q\right)^2+ \left(\Ga_{31}^{q*}\right)^2\right]\left(\frac{m_{B_d}}{m_b+m_d} \right)^2\right\} \left(  \fr{g^2}{m^2_{\mathcal{H}_1}}-
\frac{g^2}{m^2_\mathcal{A}} \right) m_{B_d} f_{B_d}^2 \nonumber \crn
&& - \frac{1}{4}\mathrm{Re}\left\{ \Ga_{31}^{q*}\Ga_{13}^q  \left(\frac{1}{6}+\frac{m_{B_d}^2}{\left(m_b+m_d\right)^2} \right)\right\}\left(\frac{g^2}{m^2_{\mathcal{H}_1}} +\frac{g^2}{m_{\mathcal{A}}^2}\right) m_{B_d} f_{B_d}^2,
\label{ptdt17}\eea

\bea
\Delta m_{B_s}^{\text{NP}} &\simeq& \frac{2}{3} \mathrm{Re}\left\{ L^2_{23}-\left[\fr 3 2+ \left(\fr{m_{B_s}}{m_s+m_b}\right)^2\right]L_{23}R_{23}+R^2_{23}\right\}\fr{g^2}{m^2_{Z_2}}m_{B_s} f^2_{B_s}\nonumber \crn
&& +\frac{5}{48}\mathrm{Re} \left\{\left[ \left(\Ga_{23}^q\right)^2+ \left(\Ga_{32}^{q*}\right)^2\right]\left(\frac{m_{B_s}}{m_b+m_s} \right)^2\right\} \left(  \fr{g^2}{m^2_{\mathcal{H}_1}}-
\frac{g^2}{m^2_\mathcal{A}} \right) m_{B_s} f_{B_s}^2\nonumber \crn
&& - \frac{1}{4}\mathrm{Re}\left\{ \Ga_{32}^{q*}\Ga_{23}^q  \left(\frac{1}{6}+\frac{m_{B_d}^2}{\left(m_b+m_s\right)^2} \right)\right\}\left(\frac{g^2}{m^2_{\mathcal{H}_1}} +\frac{g^2}{m_{\mathcal{A}}^2}\right) m_{B_s} f_{B_s}^2.
\label{ptdt18}\eea
The hadronic matrix elements have been determined by using PCAC \cite{Gabbiani:1996hi}. The total contribution to the mass difference in
meson mixing systems arises from both standard model (SM) and new physics (NP) effects. This can be expressed as:
\bea 
\Delta m_{K,B_d,B_s}=\Delta m^{\text{SM}}_{K,B_d,B_s}+\Delta m^{\text{NP}}_{K,B_d,B_s}. \label{total}
\eea 
The SM predictions and experimental values for these meson mass differences is provided in the Table \ref{SM_exp data} 
\begin{table}[H]
	\protect\caption{\label{SM_exp data} The SM predictions and corresponding world average experimental values for flavor-changing observables}
	\begin{centering}
		\begin{tabular}{|c|c|c|}
			\hline
			Observables & SM predictions  & Experimental values  \tabularnewline
			\hline 
			$\Delta m_K$ & $0.467\times 10^{-2}   \  \text{ps}^{-1}$ \cite{Workman:2022ynf}& $0.5293(9)\times 10^{-2}  \  \text{ps}^{-1}$ (PDG) \cite{Workman:2022ynf} \tabularnewline
			$\Delta m_{B_s}$ & $18.77(86) \  \text{ps}^{-1}$\cite{Lenz:2019lvd} &  $17.765(6)  \  \text{ps}^{-1}$  (HFLAV) \cite{HFLAV:2022pwe} \tabularnewline
			$\Delta m_{B_d}$ & $0.543(29)\ \text{ps}^{-1} $ \cite{Lenz:2019lvd}&	$0.5065(19)  \  \text{ps}^{-1}s$ (HFLAV) \cite{HFLAV:2022pwe}\tabularnewline
			$\text{BR}(B_s\to \mu^+\mu^-)$ & $(3.66 \pm 0.14 ) \times 10^{-9} $ \cite{Beneke:2019slt} &	$(3.45\pm 0.29) \times 10^{-9}$  (HFLAV) \cite{LHCb:2021awg} \tabularnewline 
			$\text{BR}(\bar{B}\to X_s \gamma)$ & $(3.40 \pm 0.17) \times 10^{-4}$\cite{Misiak:2020vlo} &	$(3.49\pm 0.19) \times 10^{-4}$ (HFLAV) \cite{HFLAV:2022pwe}  \tabularnewline	
			\hline 
		\end{tabular}
		\par
	\end{centering}	
\end{table}
In $K^0-\bar{K}^0$ system, the uncertainties are quite large because the lattice QCD calculations for long-distance effect are not well controlled. Therefore, we assume the predicted theory contributes about 30\% to $\Delta m_K$, it reads    
\bea 
\fr{(\Delta m_{K})_{\text{SM}}}{(\Delta m_{K})_{\text{exp}}}=1(1\pm0.3), \eea 
and translates to the following constraint
\bea  \fr{(\Delta m_{K})_{\text{NP}}}{(\Delta m_{K})_{\text{exp}}}\in [-0.3,0.3]. 
\label{constraint1}\eea
The SM contributions for $\Delta m_{B_s,B_d}$ are more accurate compared with $\Delta m_K$, we have the following constraints, as can be seen from Table .\ref{SM_exp data}. We combine in quadrature the relative uncertainties in both SM and experiment and get the $2\sigma$ following constraints   
\bea \fr{(\Delta m_{B_d})_{\text{NP}}}{(\Delta m_{B_d})_{\mathrm{Exp}}}\in[-0.187,0.043],   \hs \fr{(\Delta m_{B_s})_{\text{NP}}}{(\Delta m_{B_s})_{\mathrm{Exp}}}\in[-0.153,0.04] . \label{constraint2}  \eea 	
\subsubsection{$B_s \to \mu^+ \mu^-, B\to K^* \mu^+ \mu^-$,  $  B^+ \to K^+ \mu^+ \mu^-$}
The B-meson decay rates are extremely sensitive to new physics. Quark FCNCs and lepton currents determine the effective Hamiltonian for the meson decay processes: $ B_s \to \mu+ \mu-, B \to K* \mu+ \mu-$, and $B^+ \to K^+ \mu^+ \mu^-$. 
The interaction terms of leptons with the new scalar fields, $\mathcal{H}_1, \mathcal{A}$ are obtained from the Yukawa Lagrangian (\ref{Yuka1}) as follows
\bea
\mathcal{L}_{l-\mathcal{H}_1/\mathcal{A}}=-t_\al\frac{g}{2m_W} \bar{l}_{aL} m_{l_a} l_{aR}\left( \mathcal{H}_1+i \mathcal{A}\right)+H.c.,
\eea 
and the new neutral gauge boson couples with two charged leptons, such as
\bea
\mathcal{L}_{l-Z_2}= -\fr{g}{2c_W}\bar{l}\ga^\mu \left(g_V^{Z_2}(f)-g_A^{Z_2}(f) \ga_5\right) l Z_{2 \mu},
\eea
where $g_V^{Z_2}(f), g_A^{Z_2}(f) $ can be found in  \cite{VanDong:2022cin}.
Combining with the FCNCs given in Eqs.(\ref{fcnc_higgs1},\ref{fcnczp}), we obtain the effective Hamiltonian as follows
\bea
H_{\text{eff}}= -\frac{4 G_F}{\sqrt{2}}V_{tb} V_{ts}^* \sum_{j=9,10,9',10',S,P} \left(C_j(\mu)\mathcal{O}_J (\mu)+C_j^\prime(\mu)\mathcal{O}^\prime_J (\mu)\right)
\eea
with 
\bea
\mathcal{O}_9 && = \frac{e^2}{\left(4\pi \right)^2} \left(\bar{s}\ga_\mu P_L b \right) \left(\bar{l}\ga^\mu l\right), \hs 
\mathcal{O}^\prime_9= \frac{e^2}{\left(4\pi \right)^2} \left(\bar{s}\ga_\mu P_R b \right) \left(\bar{l}\ga^\mu l\right), \nonumber  \crn \mathcal{O}_{10} && = \frac{e^2}{\left(4\pi \right)^2} \left(\bar{s}\ga_\mu P_L b \right) \left(\bar{l}\ga^\mu \ga^5 l\right), \hs 
\mathcal{O}_{10}^\prime= \frac{e^2}{\left(4\pi \right)^2} \left(\bar{s}\ga_\mu P_R b \right) \left(\bar{l}\ga^\mu \ga^5 l\right), \nonumber \crn
\mathcal{O}_S && = \frac{e^2}{(4\pi)^2}\left(\bar{s}P_R b\right) \left( \bar{l}l\right), \hs \mathcal{O}^\prime_S =\frac{e^2}{(4\pi)^2}\left(\bar{s}P_L b\right) \left(\bar{l} l\right), \nonumber \crn
\mathcal{O}_P && = \frac{e^2}{(4\pi)^2}\left(\bar{s}P_R b\right) \left( \bar{l} \ga_5l\right), \hs \mathcal{O}^\prime_P =\frac{e^2}{(4\pi)^2}\left(\bar{s}P_L b\right) \left(\bar{l}\ga_5 l\right).
\eea
The Wilson coefficients (WCs), $C_i,C^\prime_i$, are dived into 
two parts, $C_i^{(\prime)}=C_i^{(\prime) \text{SM}}+C_i^{(\prime)\text{NP}}$, where only the $C_{9,10}^{\text{SM}}$ are non vanished and their central points are given in \cite{Beneke:2017vpq}, $C_9^{\text{SM}}=4.344, C_{10}^{\text{SM}}=-4.198$. The contributions of NP to the WCs are
\bea
C_9^{\text{NP}} && =L_{23}\frac{m_W^2}{c_W V_{tb}V^*_{ts}}\fr{(4\pi)^2}{e^2}\fr{ g_V^{Z_2}(\mu)}{m_{Z_2}^2}, \hs C_9^{\prime \text{NP}}=R_{23}\frac{m_W^2}{c_W V_{tb}V^*_{ts}}\fr{(4\pi)^2}{e^2}\fr{ g_V^{Z_2}(\mu)}{m_{Z_2}^2}, \nonumber \crn
C_{10}^{\text{NP}}&& =-L_{23}\frac{m_W^2}{c_W V_{tb}V^*_{ts}}\fr{(4\pi)^2}{e^2}\fr{ g_A^{Z_2}(\mu)}{m_{Z_2}^2}, \hs C_{10}^{\prime \text{NP}}=-R_{23}\frac{m_W^2}{c_W V_{tb}V^*_{ts}}\fr{(4\pi)^2}{e^2}\fr{ g_A^{Z_2}(\mu)}{m_{Z_2}^2}, \nonumber 
\crn
C_{\text{S}}^{\text{NP}} && =\fr{8\pi^2}{e^2}\fr{1}{V_{tb} V^*_{ts}}\frac{\Ga^q_{23}}{m_{\mathcal{H}_1}^2}t_\al m_{l}, \hs \hs \hs \hs C_{\text{S}}^{\prime \text{NP}} =\fr{8\pi^2}{e^2}\fr{1}{V_{tb} V^*_{ts}}\frac{\left(\Ga^q_{32}\right)^*}{m_{\mathcal{H}_1}^2}t_\al m_{l},\nonumber 
\crn
C_{\text{P}}^{\text{NP}} && =-\fr{8\pi^2}{e^2}\fr{1}{V_{tb} V^*_{ts}}\frac{\Ga^q_{23}}{m_{\mathcal{A}}^2}t_\al m_{l}, \hs \hs \hs \hs C_{\text{P}}^{\prime \text{NP}} =-\fr{8\pi^2}{e^2}\fr{1}{V_{tb} V^*_{ts}}\frac{\left(\Ga^q_{32}\right)^*}{m_{\mathcal{A}}^2}t_\al m_{l}. \label{WCs}
\eea
Theoretically the branching ratio of the $B_s \to l_\al^+ l_\al^-$ decay is determined by
\bea
&&\mathrm{BR}(B_s \to l_\al^+ l_\al^-)=\frac{\tau_{B_s}}{64 \pi^3}\al^2 G_F^2 f_{B_s}^2|V_{tb}V^*_{ts}|^2m_{B_s}\sqrt{1-\fr{4m_{l_\al}^2}{m^2_{B_s}}} \nonumber \crn 
&&\times \left\{ \left( 1-\fr{4m^2_{l_\al}}{m^2_{B_s}}\right)\left|\fr{m_{B_s}^2}{m_b+m_s}\left(C_\text{S}-C_\text{S}^\prime \right) \right|^2 + \left|2m_{l_\al}\left(C_{10}-C^\prime_{10}\right)+\fr{m_{B_s}^2}{m_b+m_s}\left( C_\text{P}-C_\text{P}^\prime\right) \right|^2\right\}
\eea
with $\tau_{B_s}$ is a lifetime of the $B_s$. Because of the effect of oscillations of the meson, the experimental results
relate to theories presented in \cite{DeBruyn:2012wj}:
\bea
\mathrm{BR}(B_s \to l_\al^+ l_\al^-)_{\text{exp}}\sim \fr{1}{1-y_s}\mathrm{BR}(B_s \to l_\al^+ l_\al^-)_{\text{theory}}
\eea
with $y_s=\frac{\Delta \Ga_{B_s}}{2 \Ga_{B_s}}$ and is numerically given in the Table. \ref{input-par}.

We would like to emphasize that the SM predicted the following outcomes \cite{Bobeth:2013uxa}\cite{Beneke:2019slt}:  
\bea
\mathrm{BR}(B_s \to e^+e^-)_{\text{SM}}&& =\left(8.54 \pm 0.55\right) \times 10^{-14}, \nonumber \crn  \mathrm{BR}(B_s \to \mu^+\mu^-)_{\text{SM}}&& =\left(3.66 \pm 0.14 \right) \times 10^{-9}.
\eea
While the experimental bounds have been given in \cite{LHCb:2020pcv} as follows:
\bea
\mathrm{BR}(B_s \to e^+e^-)_{\text{exp}}&& < 9.4 \times 10^{-9},\eea
and BR$(B_s \to \mu^+\mu^-)$ has the most current average experimental value given in Table \ref{SM_exp data} which is benefited from the newest results of LHCb \cite{LHCb:2021awg}, and CMS \cite{CMS:2022mgd}. This upgrade of  BR$(B_s \to \mu^+\mu^-)$ bring the measurement and SM prediction closer, and therefore the NP contribution, if having will be very small.  

Similarly the meson mixing systems, we combine both uncertainties from SM and experimental of BR$(B_s\to \mu^+\mu^-)$ and obtain the $2\sigma$ range as follows
\bea 
\fr{\text{BR}(B_s\to \mu^+\mu^-)_{\text{exp}}}{\text{BR}(B_s\to \mu^+\mu^-)_{\text{SM}}}&& =\fr{1}{1-y_s}\fr{\left(1-\fr{4m_{\mu}^2}{m_{B_s}^2}\right)|\tilde{S}|^2+|\tilde{P}|^2}{|C_{10}^{\text{SM}}|^2}\in [0.7684,1.1168] \label{Bsmm_constraint}
\eea 
with 
\bea 
&& \tilde{P}=(C_{10}-C_{10}')+\fr{m_{B_s}^2}{2m_{\mu}(m_b+m_s)}(C_P-C_P'), \crn 
&&  \tilde{S}=\fr{m_{B_s}^2}{2m_{\mu}(m_b+m_s)}|C_S-C_S'|^2.
\eea
\subsubsection{$\bar{B}\to X_s \gamma$}
The contributions to the decay processes, $\bar{B}\to X_s \gamma$, come from the FCNCs coupled by both new neutral gauge boson $Z^{\prime}$ and new scalars $\mathcal{H},\mathcal{A}$.  Their relevant Lagrangian are obtained from Eqs. (\ref{fcnczp},\ref{fcnc_higgs1}). The effective Hamiltonian for the decay $b \rightarrow s \gamma$ is expressed by 
\bea \mathcal{H}_{\text{eff}}^{b \rightarrow s \gamma}&&=-\fr{4G_F}{\sqrt{2}}V_{tb}V_{ts}^{*}[C_7(\mu_b)\mathcal{O}_7+C_8(\mu_b)\mathcal{O}_8 + C_7'(\mu_b)\mathcal{O}'_7+C_8'(\mu_b)\mathcal{O}'_8],\eea  
with $\mu_b=\mathcal{O}(m_b)$ is the energy scale of the decay $b\to s \gamma$. The electromagnetic and chromomagnetic dipole operators $\mathcal{O}_7,\mathcal{O}_8$ are defined as 
\bea
\mathcal{O}_7=\fr{e}{(4\pi)^2} m_b(\bar{s}_{\al}\sigma_{\mu \nu}P_R b_{\al})F^{\mu \nu}, \hs \mathcal{O}_8=\fr{g_s}{(4\pi)^2} m_b(\bar{s}_{\al}\sigma_{\mu \nu}T^a_{\al \beta}P_R b_{\beta})G^{a \mu \nu},
\eea
and the primed operators $\mathcal{O}_{7,8}'$ are obtained by replacing $P_L \leftrightarrow P_R$. The primed Wilson coefficients (WCs) $C_{7,8}'$ are  obtained by replacing $P_L \to P_R$. It should be noted that in the limit $m_b \gg m_s \sim 0$, the WCs $C_{7,8}'$ can be ignored, and there are left unprimed WCs $C_{7,8}$.The WCs $C_{7,8}(\mu_b)$ split as the sum of the SM and NP contributions
\bea
C_{7,8}(\mu_b)=C_{7,8}^{\text{SM}}(\mu_b)+C_{7,8}^{\text{NP}}(\mu_b),  \label{bsgamma}
\eea 
with $C_{7,8}^{\text{SM}}$ are the SM WCs which are first given by \cite{Inami:1980fz}, at the scale $\mu \sim m_W$  
\bea
&& C^{\text{SM}(0)}_7(m_W)=\fr{m_t^2}{m_W^2}f_{\gamma}\left(\fr{m_t^2}{m_W^2}\right), \hs C_{8}^{\text{SM}(0)}(m_W )= \fr{m_t^2}{m_W^2}f_g \left(\fr{m_t^2}{m_W^2}\right), \eea
where the index \text{0} indicates that the WCs are calculated without QCD correction. The \text{NP} contributions to the WCs , $C_{7,8}^{\text{NP}}$, come from the charged scalar  currents given in Eq. (\ref{bs1}) and the  \text{FCNCs} given in Eqs. (\ref{fcnc_higgs1}), (\ref{fcnczp}). We can divide the contributions as follows:
\bea C_{7,8}^{\text{NP}(0)}=C_{7,8}^{H^{\pm}(0)}(m_{H^{\pm}})+C_{7,8}^{Z^{\prime}(0)}(m_{Z^{\prime}})+C_{7,8}^{\mathcal{H}_1,\mathcal{A}(0)}(m_{\mathcal{H}_1,\mathcal{A}}), \eea
where
\bea  
&&  C_{7}^{H^{\pm}(0)}(m_{H^{\pm}})=t_{\al}^2\fr{m_t^2}{m_{H^{\pm}}^2}\left[\fr{1}{3} f_{\gamma}\left(\fr{m_t^2}{m_{H^{\pm}}^2}\right)-f^{\prime}_{\gamma} \left(\fr{m_t^2}{m_{H^{\pm}}^2}\right)\right], \crn
&& C_{8}^{H^{\pm}(0)}(m_{H^{\pm}} )= t_{\al}^2\fr{m_t^2}{m_{H^{\pm}}^2}\left[\fr{1}{3} f_{g}\left(\fr{m_t^2}{m_{H^{\pm}}^2}\right)-f^{\prime}_{g} \left(\fr{m_t^2}{m_{H^{\pm}}^2}\right)\right].  \eea 
The functions $f_{\ga,g}$ and $f'_{\ga,g}$ are defined by 
\bea
&& f_{\gamma}(x)=\fr{(7-5x-8x^2)}{24(x-1)^3}+\fr{x(3x-2)}{4(x-1)^4} \ln{x}, \hs 
f^{\prime}_{\gamma}(x)=\fr{(3-5x)}{12(x-1)^2}+\fr{(3x-2)}{6(x-1)^3}\ln{x}, \crn 
&& f_{g}(x)=\fr{2+5x-x^2}{8(x-1)^3}-\fr{3x}{4(x-1)^4}\ln{x}, \hs 
f^{\prime}_{g}(x)=\fr{3-x}{4(x-1)^2}-\fr{1}{2(x-1)^3}\ln{x}.
\eea

It is important to comment that the $C_{7}^{Z^{\prime}(0)}(m_{Z^{\prime}})$ is suppressed by a factor $m_{W}^2/m_{Z'}^2\sim v^2/\La^2 \ll 1$, which is much smaller than SM and can be ignored. Similarly, for FCNC associated scalars $\mathcal{H}_1,\mathcal{A}$, their corresponding WCs are proportional with $\fr{\Ga^d_{3i}\Ga^d_{3j}}{m_{\mathcal{H},\mathcal{A}}^2}\sim \fr{m^2_d}{m_{\mathcal{H},\mathcal{A}}^2}\ll 1$, therefore we can also remove these terms in our calculation. 

The \text{QCD} corrections to $b \rightarrow s \gamma$ are necessary for the analysis. In SM, $C_{7,8}^{\text{SM}}$ were calculated up to  Next-to-Next-Leading Order (NNLO), specifically, we compute  $C_7^{\text{SM}}(\mu_b)=-0.3636$ for $\mu_b=2.0$ GeV, based on the Refs.  \cite{Misiak:2006ab,Czakon:2006ss,Czakon:2015exa}. However, the \text{NP} contributions to the $C_{7, 8}^{\text{NP}}$ have been considered at the Leading Order (\text{LO}) \cite{Buras:2012dp}, \cite{Buras:2011zb}. In this work, we study the effect of \text{QCD} corrections on the $C_{7,8}^{\text{NP}}$ at the \text{LO}. If including the \text{LO} of \text{QCD} corrections, $C_7^{H^{\pm}}$ at the scale $\mu_b$ has the forms as \cite{Buras:2012dp}, \cite{Buras:2011zb} 
\bea 
&& C_{7}^{H^{\pm}}(\mu_b)  =\ka_7C_7^{H^{\pm}}(m_{H^{\pm}})+\ka_8 C_8^{H^{\pm}}(m_{H^{\pm}}),\label{WCs2}\eea 
where $\ka_{7,8}$ are so called "magic numbers" and given in  \cite{Buras:2011zb}.The branching ratio for the considering decay is given as \cite{Buras:2011zb}
\bea
\text{BR}(\bar{B}\to X_s\ga)&=&\fr{6\al_{\text{em}}}{\pi C}\left|\fr{V_{ts}^*V_{tb}}{V_{cb}}\right|^2 [|C_7(\mu_b)|^2+N(E_{\gamma})]\text{BR}(\bar{B}\rightarrow X_c l\bar{\nu}) , \label{bra1}
\eea 
where $N(E_{\gamma})$ is a non-perturbative contribution which amounts around $4\%$ of the branching ratio. We compute the leading order contribution to $N(E_{\gamma})$ followed the Eq. (3.8) in Ref. \cite{Misiak:2020vlo} and obtain $N(E_{\ga})\simeq 3.3\times 10^{-3}$. $C$ is the semileptonic phase-space factor $C=|V_{ub}/V_{cb}|^2\Ga(\bar{B}\rightarrow X_c e\bar{\nu}_e)/\Ga(\bar{B}\rightarrow X_u e\bar{\nu}_e)$  and BR$(\bar{B}\rightarrow X_c l\bar{\nu})$ is the branching ratio for semi-leptonic decay . 

To combine both the SM and experimental uncertainties from Table \ref{SM_exp data} and reduce the amount of input parameters for this observable, we consider the ratio between SM and world average experimental value, thus obtaining the following $2\sigma$ constraint  
\bea
\fr{|C_7^{\text{NP}}|^2+2C_7^{\text{SM}}\text{Re}[C_7^{\text{NP}}]}{|C_7^{\text{SM}}|^2+N(E_{\ga})}\in [-0.1252,0.1782] \label{Bsg_constraint} \eea 
\subsubsection{Radiative decays of top quark $t\to u(c)\gamma$}
Similar to the down-type quark sector, the model also features several flavor observables related to up-type quark sector, including branching ratio of radiative top quark decays $t\to u(c)
\gamma$. These processes can be generated at one-loop level by FCNC interactions associated with $Z'$ gauge boson in Eq. (\ref{fcnczp}) and  by interactions of charged Higgs bosons $H^{\pm}$ in Eq. (\ref{bs1}). 

The branching ratio of radiative top quark decays $t\to u(c)
\gamma$ is given by
\bea
\text{BR}(t\to u_i\ga)&=&\fr{\Ga(t\to u(c)\ga)}{\Ga^{\text{total}}_{t}}=\fr{m_t^3(|C^{tu_i\ga}_L|^2+|C_R^{tu_i\ga}|^2)}{16\pi\Ga^{\text{total}}_t},\hs u_i=u,c
\eea 
where $C^t_{L,R}$ are the coefficients obtained in the limit $m_t\gg m_u, m_c$. They can be split into different contributions as follows :
\bea 
&& C^{tu_i,Z'}_L=\fr{iem_t}{16\pi^2m_{Z'}^2}\sum_{i=1}^3\left[R^*_{iI}R_{i3}h\left(\fr{m_{u_i}^2}{m_{Z'}^2}\right)+\fr{m_{u_i}}{m_t}R^*_{iI}L_{i3}h'\left(\fr{m_{u_i}^2}{m_{Z'}^2}\right)\right], \crn 
&& C^{tu_i,Z'}_R=\fr{iem_t}{16\pi^2m_{Z'}^2}\sum_{i=1}^3\left[L^*_{iI}L_{i3}h\left(\fr{m_{u_i}^2}{m_{Z'}^2}\right)+\fr{m_{u_i}}{m_t}L^*_{iI}R_{i3}h'\left(\fr{m_{u_i}^2}{m_{Z'}^2}\right)\right],\crn 
&& C^{tu_i,\mathcal{H}^{\pm}}_L=\fr{iem_t}{16\pi^2m_{\mathcal{H}^{\pm}}^2}\fr{g^2}{2m_W^2}\sum_{i=1}^3\left[\mathcal{X}^*_{iI}\mathcal{X}_{i3}g\left(\fr{m_{d_i}^2}{m_{\mathcal{H}^{\pm}}^2}\right)+\fr{m_{d_i}}{m_t}\mathcal{X}^*_{iI}\mathcal{Y}_{i3}g'\left(\fr{m_{d_i}^2}{m_{\mathcal{H}^{\pm}}^2}\right)\right],\crn 
&& C^{tu_i,\mathcal{H}^{\pm}}_R=\fr{iem_t}{16\pi^2m_{\mathcal{H}^{\pm}}^2}\fr{g^2}{2m_W^2}\sum_{i=1}^3\left[\mathcal{Y}^*_{iI}\mathcal{Y}_{i3}g\left(\fr{m_{d_i}^2}{m_{\mathcal{H}^{\pm}}^2}\right)+\fr{m_{d_i}}{m_t}\mathcal{Y}^*_{iI}\mathcal{X}_{i3}g'\left(\fr{m_{d_i}^2}{m_{H^{\pm}}^2}\right)\right].
\eea 
 The loop functions $h^{(')}(x),g^{(')}(x)$ are defined by
\bea 
&& h(x)=\fr{8-30x+9x^2-5x^3}{18(x-1)^3}+\fr{x^2}{(x-1)^4}\ln{x},\crn  
&& h'(x)=\fr{x^2+x+4}{3(x-1)^2}-\fr{2x}{(x-1)^3}\ln{x},\crn 
&& g(x)=\fr{5-10x-7x^2}{36(x-1)^3}+\fr{x(3x-1){6(x-1)^4}}\ln{x},\crn
&& g'(x)=-\fr{2}{3(x-1)^2}+\fr{3x-1}{3(x-1)^3}\ln{x}
\eea 
The predicted branching ratios of top quark decays have to be compared the upper experimental limits \cite{Workman:2022ynf}:  BR$(t\to u\gamma)_{\text{exp}}<0.85\times 10^{-5}$ and  BR$(t\to c\gamma)_{\text{exp}}<4.2\times 10^{-5}.$ 
\subsection{Lepton flavor phenomenology}
In this model, charged Higgs bosons $\mathcal{H}^\pm$ and right-handed neutrinos $\nu_{aR}$ contribute to lepton flavor violation (LFV) processes at the one-loop level. This includes branching ratios of radiative decays $(e_I)\to e_J \gamma$, three-body leptonic decays $e_I\to 3e_J$ ($I\neq J$) and anomalous magnetic moments for electron (muon) $\Delta a_{e,mu}$. The effective Hamiltonian descring these observables is given by
\bea
\mathcal{H}^{\text{lepton}}_{\text{eff}}=C^{JI}_L\bar{e}_J\sigma_{\mu\nu}P_L e_IF^{\mu\nu}+(L\to R), \label{leptonfcnc}\eea 
where the coefficients $C^{JI}_{L,R}$ are obtained by  one-loop diagram calculations. In the limit  $m_{e_{I}}\gg m_{e_J}$,  these coefficients can be expressed as: 
\bea 
C^{JI}_{L,R}=C^{JI,W^{\pm}}_{L,R}+C^{JI,\mathcal{H}^{\pm}}_{L,R},
\eea 
with  
\bea 
&& C^{JI,W^{\pm}}_R=\fr{iem_I}{16\pi^2m_W^2}\sum_{i=1}^3(V^{\nu *}_L)_{Ii}(V^{\nu}_L)_{Ji}r\left(\fr{m_{\nu_{iL}^2}}{m_W^2}\right), \hs C^{IJ,W^{\pm}}_L=0,\crn 
&&
C^{JI,\mathcal{H}^{\pm}}_R=\fr{iem_Is_{\al}^2}{16\pi^2m_{\mathcal{H}^{\pm}}^2}\sum_{i=1}^3h^{\nu*}_{Ii}h^{\nu}_{Ji}k\left(\fr{m_{\nu_{iR}}^2}{m_{\mathcal{H}^{\pm}}^2}\right), \hs C^{IJ,\mathcal{H}^{\pm}}_L=0, 
\eea 
where $C^{Ji,W^{\pm}}_{L,R}$ are contributions from the SM $W^{\pm}$ boson, and $r(x)$ and $k(x)$ are loop functions which are given as follows: 
\bea 
&& r(x)=\fr{10+x[-33+(45-4x)x]}{12(x-1)^3}-\fr{3x^3}{2(x-1)^4}\ln{x},\crn  
&& k(x)=\fr{-1+5x+2x^2}{12(x-1)^3}-\fr{x^2}{2(x-1)^4}\ln{x}. 
\eea 
The branching ratios of the LFV decay processes, $e_I\to e_J \gamma$, are determined by 
\be \text{BR}(e_I\to e_J \gamma)=\fr{m_{e_I}^3}{4\Ga_{e_I}}(|C^{JI}_L|^2+|C^{JI}_R|^2), \ee
where $\Ga_{e_I}$ is the total decay width of decaying lepton $e_I$. 
The electron and muon anomalous magnetic moments $\Delta a_{e,\mu}$ read
\bea \Delta a_{e_I} &=&-\fr{4m_{e_I}}{e}\text{Re}[C_R^{II}],\label{lfc1} \crn
C^{II,\mathcal{H}^{\pm}}_R&=&\fr{iem_Is_{\al}^2}{16\pi^2m_{H^{\pm}}^2}\sum_{i=1}^3|h^{\nu}_{Ii}|^2r\left(\fr{m_{\nu_{iR}}^2}{m_{\mathcal{H}^{\pm}}^2}\right) 
\eea 
It is important to note that the leptonic observables like BR$(e_I\to e_J \gamma)$  are proportional to the squared product of two Yukawa couplings $h^{\nu *}_{Ii}h^{\nu}_{Ji}$. The small neutrino mass  $m_{\nu}\sim \mathcal{O}(0.1)$ eV,  obtained from type-I seesaw mechanism, implied that these Yukawa couplings are highly constrained, $h^{\nu}\sim 10^{-5}$. Consequently,  BR$(e_I\to e_J \gamma)\sim |h^{\nu *}_{Ii}h^{\nu}_{Ji}|^2\sim 10^{-20}$,  combined with the overall factor $\fr{iem_I}{16\pi^2m_{H^{\pm}}^2}\sim 10^{-6}$, results in observables that are significantly smaller than  current upper experimental limits \cite{Workman:2022ynf} of $ \mathcal{O}(10^{-8}-10^{-13}).$  

Due to these suppressed rates, we will focus on quark flavor changing processes in the following numerical study, neglecting lepton flavor observable such as the predicted anomalous magnetic moments for the electron and muon, $\Delta a_{e,\mu}$, which are also significantly lower than measurements \cite{Muong-2:2023cdq}, \cite{Workman:2022ynf}.


\section{\label{m4} Numerical analysis}
Let us first discuss our assumptions for  input parameters. For processes related to down-type quarks sector,  we assume the left-handed quark mixing matrix $V_{dL}$ to be the CKM matrix which has been extensively measured \cite{Workman:2022ynf}. For the right-handed down type quark mixing matrix $V_{dR}$, we parameterize it as CKM matrix with three parameters $s_{12,13,23}^R$ and one CP violation phase $\delta^R$. 

In the lepton sector, we set the mixing matrix $V_{e_L}=1$, implying that the mixing matrix of active neutrinos $V_{\nu_L}$ is identified as PMNS matrix $V_{\nu_L}=V_{\text{PMNS}}$.  

For the matrix $V_{d_R}$, we consider the normal relation (NR) scenario, where $s_{12}^R> s_{23}^R> s_{13}^R$, and: 
\bea \fr{s_{13}^R}{s_{12}^R}&=&\fr{s_{13}^{\text{CKM}}}{s_{12}^{\text{CKM}}}=A\la^2+
\fr{1}{2}A\la^4\sqrt{\bar{\rho}^2+\bar{\eta}^2}\simeq 0.0157 ,\crn  \fr{s_{23}^R}{s_{12}^R}&=&\fr{s_{23}^{\text{CKM}}}{s_{12}^{\text{CKM}}}=A\la \simeq 0.1826, \eea
where $\la,A,\bar{\rho},\bar{\eta}$ are listed in the Table. \ref{commonSMvals}. 
We  also explore the inverted relation (IR) scenario, where $s_{12}^R >s_{13}^R>s_{23}^R$, and:
\bea \fr{s_{13}^R}{s_{12}^R}=\fr{s_{23}^{\text{CKM}}}{s_{12}^{\text{CKM}}}\simeq 0.1826 ,\hs \fr{s_{23}^R}{s_{12}^R}=\fr{s_{13}^{\text{CKM}}}{s_{12}^{\text{CKM}}}\simeq 0.0157.\eea
The CP violation phase is set in the range $\delta^R\in[0,2\pi]$. For the Higgs coupling $\mu_4$, we apply the condition $-\mu_4 \sim \La \gg v_1,v_2$, due to the diagonalization of the mass mixing matrix $M_S^2$. 

For up-type quark flavor processes, we assume the mixing matrices of left and right-handed up-type quarks $V_{uL,R}$ to be the  identity matrices. With theses assumptions, there are no $Z'$ contributions to flavor-changing observables related to up-type quark sectors, including branching ratios of top quark decays $t\to u(c)\gamma$, or meson oscillation $\bar{D}^0-D^0$. Therefore, the only  new contributions to up-type quark flavor changing  processes come from charged Higgs boson $H^{\pm}$. For the remaining parameters, please refer of Table \ref{input-par} and  Table \ref{commonSMvals} for their numerical values.  

For x-charge, we consider specific values $x=\pm 1/2$ and $x=\pm 1/6$. Regarding the new physics scale, we focus on the region allowed by electroweak precision tests based on both CMS and CDF measurements. 

Based on aforementioned assumptions, we will focus on  following observables: meson mass differences $\Delta m_{K},\Delta m_{B_s}, \Delta m_{B_d}$, and the branching ratios,  BR$(B_s\to \mu^+\mu^-)$, BR$(\bar{B}\to X_s\gamma)$, and BR$(t\to u(c)\gamma)$.  The $b\to sl^+l^-$ observables namely BR$(B_s\to \mu^+\mu^-)$, BR$(\bar{B}\to X_s\gamma)$ are also called as  clean observables since they have controllable theoretical uncertainties. It is important to note that new physics contributions  not only effect these clean observables
but also influence other $b\to sll$ observables such as branching ratios BR$(B^{0,+}\to K^{0,+}\mu^+\mu^-)$, Br$( B_s\to\phi \mu^+\mu^-)$, angular distributions in decays $B^+\to K^{*+}\mu^+\mu^-,B_s\to \phi \mu^+\mu^-, \La_b\to \La \mu^+\mu^-$, etc. These observables are strongly influenced by short-distance effects,  including  form factor determination and charm-loop contributions $\bar{c}c$, which are challenging to quantify accurately. This leads to substantial theoretical uncertainties compared to clean observables.
Therefore, in addition to study clean observables, we aim to assess whether the model can explain  other $b\to sl^+l^-$ observables  by comparing its predicted NP WC with constraints from global fits \cite{Alguero:2023jeh}.
\begin{table}[h]
	\setlength{\abovecaptionskip}{1pt}
	\setlength{\belowcaptionskip}{0pt}
	\protect\caption{\label{input-par} The numerical values of input parameters.}
	\begin{centering}
		\begin{tabular}{|c|c|c|c|}
			\hline
			Input parameters & Values  & Input parameters & Values \tabularnewline
			\hline 
			$ f_{K}$ & $155.7(3) \ \text{MeV}$  \cite{FlavourLatticeAveragingGroupFLAG:2021npn} &  $m_{K}$ & $497.611(13)  \ \text{MeV} $ \cite{Workman:2022ynf}
			\tabularnewline
			$f_{B_s}$ & $230.3(1.3) \ \text{MeV}$ \cite{FlavourLatticeAveragingGroupFLAG:2021npn}	 &  $ m_{B_s}$ & $5366.88(11) \ \text{MeV}$  \cite{Workman:2022ynf}\tabularnewline
			$ f_{B_d}$ & $190(1.3) \ \text{MeV}$  \cite{FlavourLatticeAveragingGroupFLAG:2021npn} &  $m_{B_d}$ & $5279.65(12)  \ \text{MeV} $ \cite{Workman:2022ynf} \tabularnewline
			$ m_{u}$ & $2.14(8) \ \text{MeV}$  \cite{FlavourLatticeAveragingGroupFLAG:2021npn} &  $m_{d}$ & $4.70(5)  \ \text{MeV} $ \cite{FlavourLatticeAveragingGroupFLAG:2021npn}\tabularnewline
			
			$\bar{m}_{c} \ (3 \ \text{GeV})$ & $0.988(11)  \ \text{GeV} $ \cite{FlavourLatticeAveragingGroupFLAG:2021npn} & $ m_{s}$ & $93.40(57) \ \text{MeV}$  \cite{FlavourLatticeAveragingGroupFLAG:2021npn} \tabularnewline
			$m_{t,\text{pole}}$ & $173.21(51)(71) \  \text{GeV} $ \cite{Czakon:2015exa} & $\bar{m}_{b}(\bar{m}_b)$ & $4.203(11)  \ \text{GeV} $ \cite{Bona:2018} \tabularnewline
			$N(E_{\ga})$ & $3.3\times 10^{-3} $ \cite{Misiak:2020vlo}& $C_7^{\text{SM}}(\mu_b=2.0 \ \text{GeV})$ & $-0.3636  $\cite{Misiak:2006ab,Czakon:2006ss,Misiak:2020vlo} \tabularnewline
			$C_9^{\text{SM}}(\mu_b=5.0 \ \text{GeV})$ & $4.344  $\ \cite{Beneke:2017vpq} &	$C_{10}^{\text{SM}}(\mu_b=5.0 \ \text{GeV})$ & $-4.198  $\  \cite{Beneke:2017vpq} \tabularnewline
			$y_s$ & $0.0645(3) $ \cite{HFLAV:2022pwe}& $\la_7$ &$  0.1 $ \tabularnewline
			$\ka_7 (\mu=5 \ \text{TeV})$ & $0.408 $ \cite{Buras:2011zb}& $\ka_8$ &$  0.129 $ \tabularnewline
			\hline 
		\end{tabular}
		\par
	\end{centering}
	\end{table}
\newpage
\begin{table}[h]
	\protect\caption{\label{commonSMvals} The common SM parameters. }
	\begin{centering}
		\begin{tabular}{|c|c|}
			\hline
			SM Parameters & Values   \tabularnewline
			\hline 
			$\la $ & $0.22548^{+0.00068}_{-0.00034} $ \cite{Charles:2015gya} \tabularnewline
			$A $ & $0.810^{+0.018}_{-0.024} $ \cite{Charles:2015gya} \tabularnewline
			$\bar{\rho} $ & $0.145^{+0.0013}_{-0.007} $ \cite{Charles:2015gya} \tabularnewline
			$\bar{\eta} $ & $0.343^{+0.0011}_{-0.012} $ \cite{Charles:2015gya} \tabularnewline 
			$m_W $ & $80.385\ \text{GeV}$  \cite{Czakon:2015exa} 
			\tabularnewline
			$m_Z $ & $91.1876 \ \text{GeV}$ \cite{Czakon:2015exa} \tabularnewline
			$G_F$ & $1.166379 \times 10^{-5} \ \text{GeV}^{-2}$ \cite{Workman:2022ynf} \tabularnewline	
			$s_W^2$ & $0.2312$ \cite{Workman:2022ynf} \tabularnewline	
			\hline 
		\end{tabular}
		\par
	\end{centering}	
\end{table}
\subsection{New physics scale is limited by CDF measurement of the $W$ gauge boson mass }
The CDF measurement of the $W$ gauge boson mass  rules out the alternative $U(1)_X$ model with $x=\pm 1/6$.  Consequently, we focus on the two remaining models with $x=\pm 1/2.$
\subsubsection{The case $x=-1/2$}
For the model with $x=-1/2$, phenomenological studies \cite{VanDong:2022cin} constraint the new physics scale to $\La \sim 5$ TeV, the electroweak scale to $v_1\in[0,55]$ GeV, and the gauge coupling ratio to $t_{\theta}=\frac{g_N}{g_X}=1$. With these constraints and the input parameters from Table. \ref{input-par},  the model has three free parameters, $s_{12}^R$, $\delta^R$ and $v_1$, which are relevant for quark flavor-changing processes. 
\begin{figure}[h]
	\centering
		\begin{tabular}{cc}
		\includegraphics[width=8.0cm]{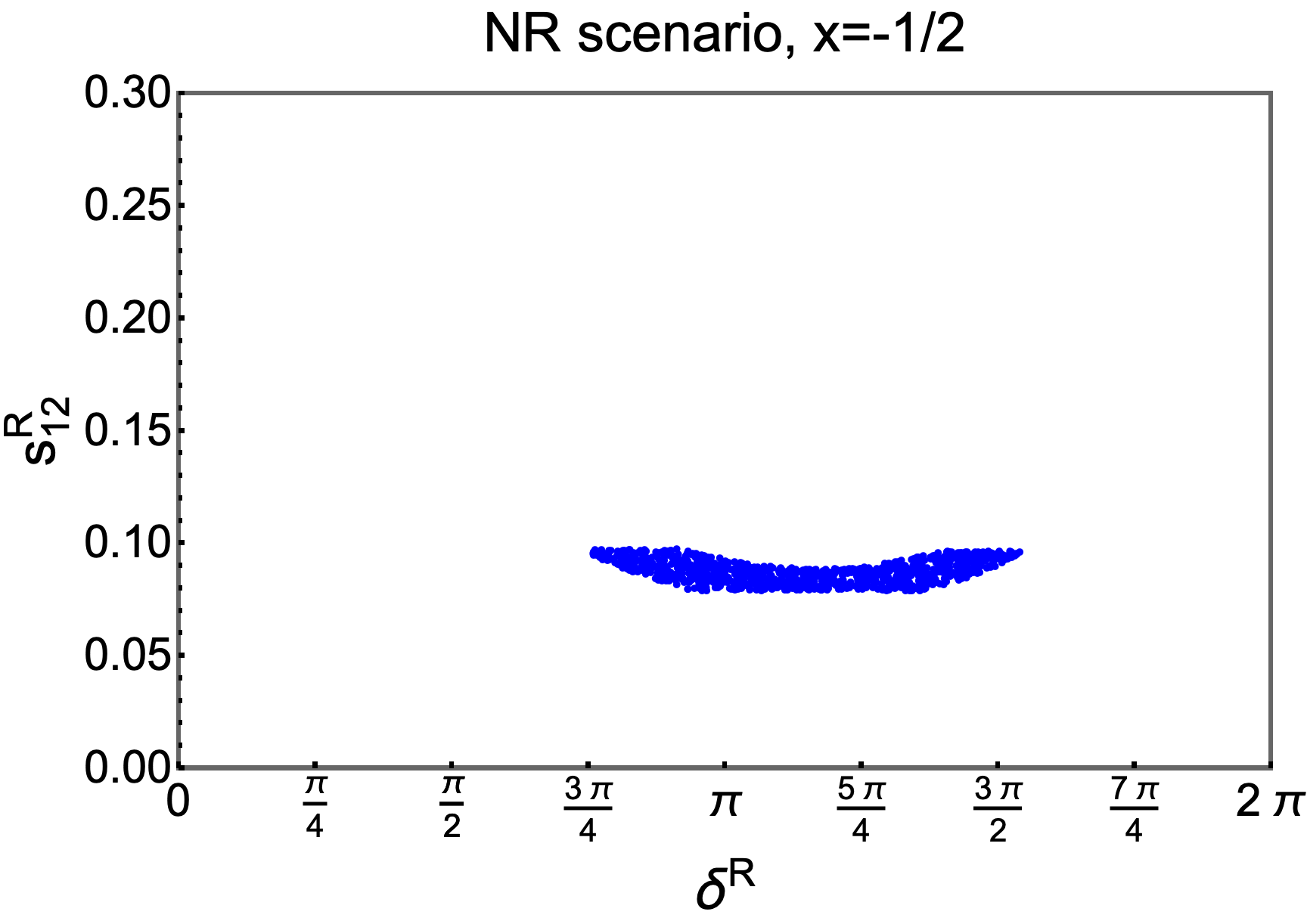}&
		\includegraphics[width=8.0cm]{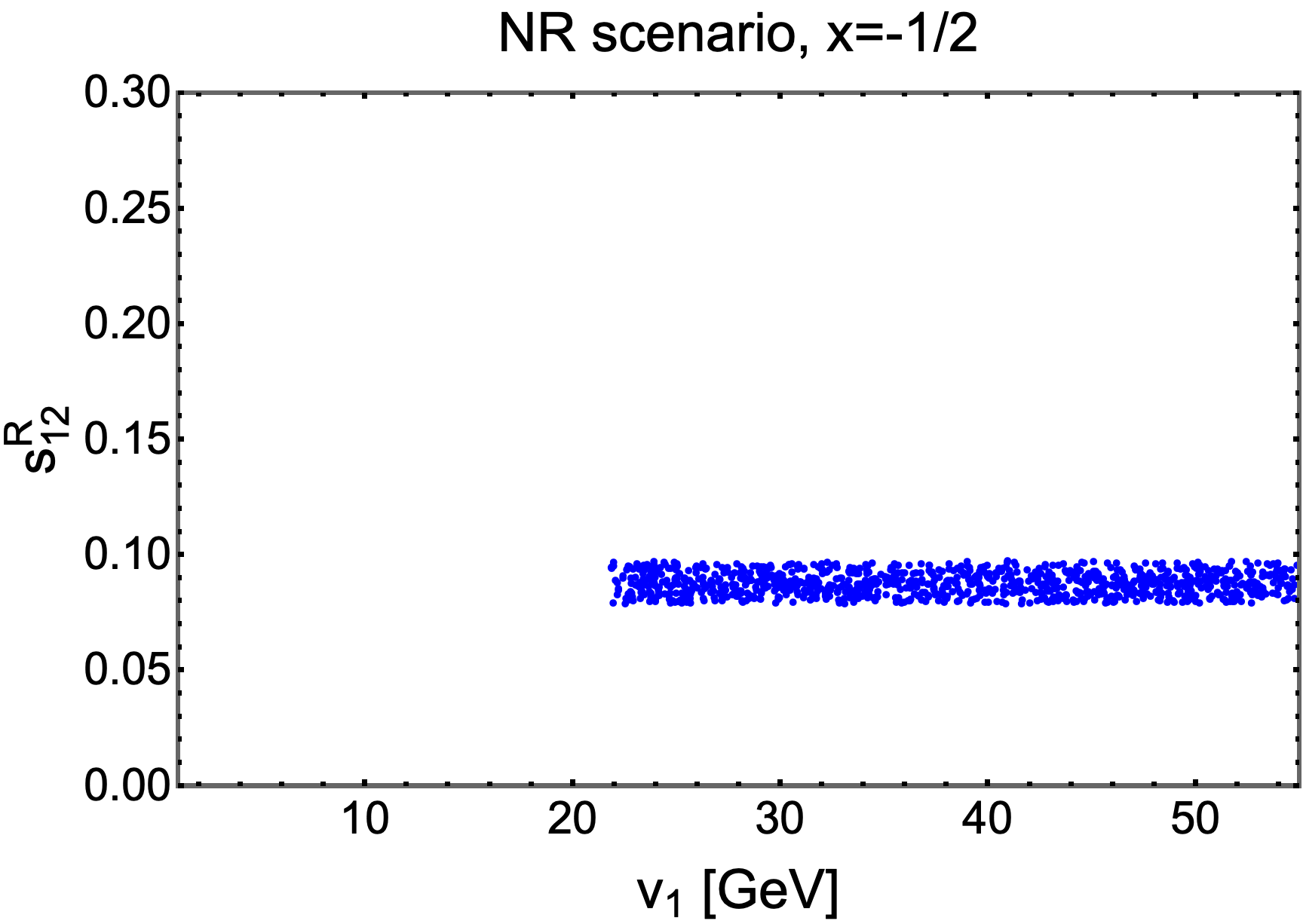}
	\end{tabular}
	\caption{ \label{case x -1/2}The left and right panels show the correlations between $s_{12}^R$ with $\delta^R$ and VEV $v_1$, respectively. The panels is plotted in the NR scenario $s_{12}^R>s_{23}^R> s_{13}^R$.}
\end{figure}
We first consider the NR scenario.  Figure. \ref{case x -1/2} shows the correlations between parameters $s_{12}^R$, $\delta^R$ and $v_1$ satisfying the constraints for flavor-changing observables: $\Delta m_{K}$  (Eq. \ref{constraint1}), $\Delta m_{B_s}$, $\Delta m_{B_d}$ (Eq.\ref{constraint2}), BR$(B_s\to \mu^+\mu^- )$ (Eq. \ref{Bsmm_constraint}), BR$(\bar{B}\to X_s \gamma)$ (Eq. \ref{Bsg_constraint}) and the upper experimental limits \cite{Workman:2022ynf}:  BR$(t\to u\gamma)_{\text{exp}}<0.85\times 10^{-5}$ and  BR$(t\to c\gamma)_{\text{exp}}<4.2\times 10^{-5}$. \\
In the left panel of  Figure. \ref{case x -1/2}, the  mixing angle $s_{12}^R$ is significantly constrained to the range $s_{12}^R\sim [0.08,0.1]$, while the CP violation phase $\delta^R$ is bounded within $[0.76\pi,1.54\pi]$. Notably, the obtained range of $s_{12}^R$ is smaller than the corresponding CKM value $s_{12}^{\text{CKM}}\simeq 0.22548.$\\The right panel shows the correlation between VEV $v_1$ and mixing angle $s_{12}^R$. We observe that $v_1$ behaves similarly to $s_{12}^\text{R}$ in the left panel.  The allowed range for  $v_1$ is quite limited,  implying $v_1\sim[22,55]$ GeV. Therefore, the model  with $x=-1/2$ in the NR case, combined with the parameter space:
$\La\sim  5 \ \text{TeV},\hs \delta^R\sim [0.76\pi,1.54\pi], \hs v_1\sim [22,55] \ \text{GeV}, \hs s_{12}^R \sim [0.08,0.1] \label{par_space_1}$, can simultaneously fit all the constraints for meson mixing $\Delta m_{K,B_s,B_d}$, and BR$(\bar{B}\to X_s \gamma)$, BR$(B_s \to \mu^+\mu^-)$ and BR($t\to u(c)\gamma$).

\begin{figure}[h]
	\centering
		\begin{tabular}{cc}
		\includegraphics[width=11.5cm]{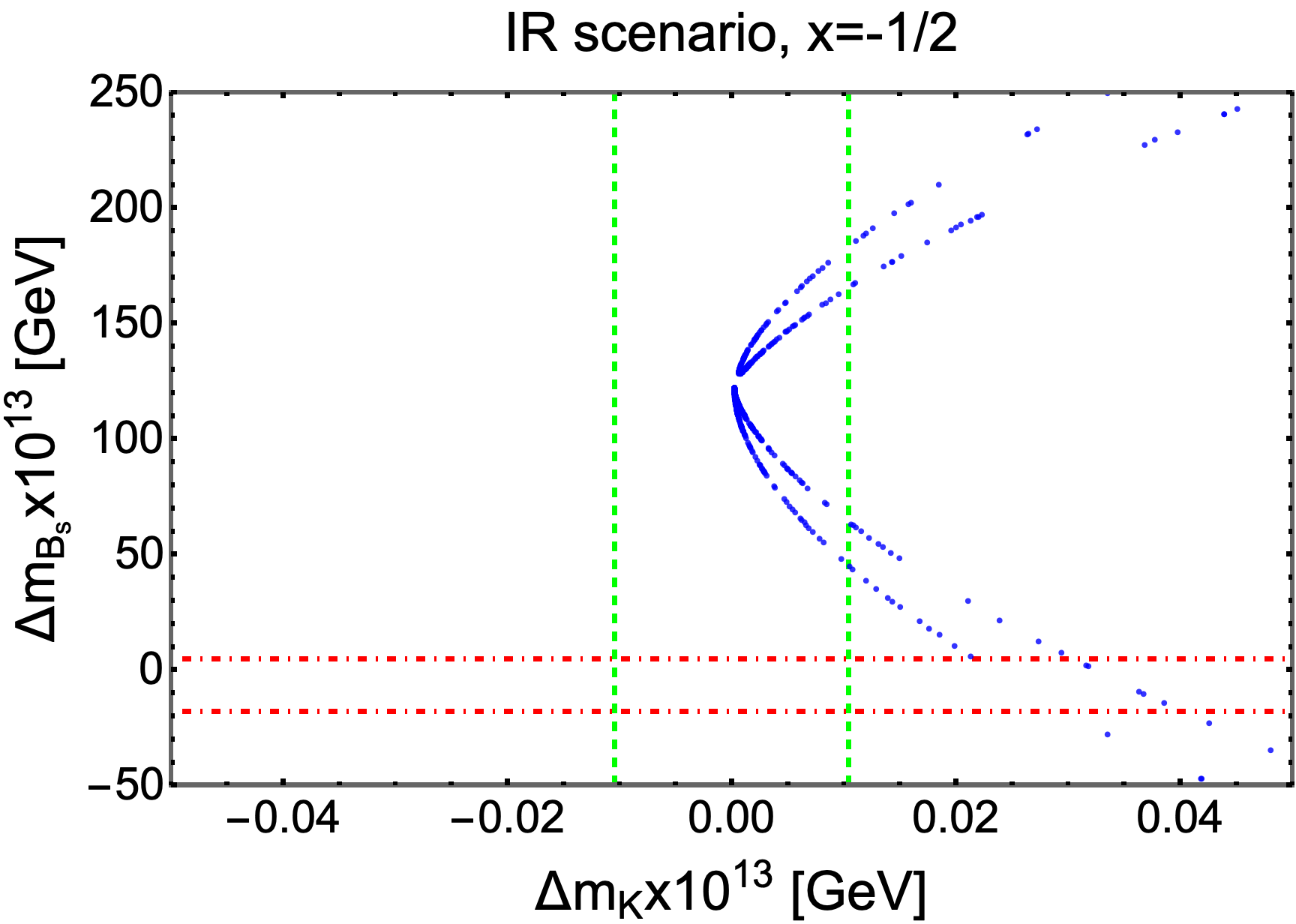}
	\end{tabular}
	\caption{ \label{xt1/2_IR} The correlation between predicted values of $\Delta m_K$ and $\Delta m_{B_s}$ for the $x=-1/2$ case. The red dot-dashed and green dashed lines present the constraints of $\Delta m_{B_s}$ and $\Delta m_{K}$ respectively, as given  by  Eq. (\ref{constraint1}) and Eq. (\ref{constraint2}).  The mixing angle parameters in $V_{dR}$ matrix are assumed to follow the IR scenario $s_{12}^R>s_{13}^R>s_{23}^R$.} 
\end{figure}
Next, we numerically study in the IR case. By randomly sampling $s_{12}^R \in [0,1] $, $\delta_R \in [0, 2\pi]$ and $v_1 \in [0,55] \text{GeV}$ while adhering to the constraints on $\Delta m_{B_d}$, BR$(\bar{B}\to X_s \gamma)$, and BR$(B_s \to \mu^+\mu^-)$, we obtain the correlation between the predicted values  for $\Delta m_K, \Delta m_{B_s}$, as depicted in Figure  \ref{xt1/2_IR}. 

The figure reveals that  points satisfying the experimental constraint for $\Delta m_{K}$ are significant more abundant than those satisfying the constraint for $\Delta m_{B_s}$. No points simultaneously satisfy both constraints. Consequently, the  IR scenario in the model with $x=-1/2$  cannot adequately explain the constraints on clean down-type quark flavor observables, unlike the NR case. This suggests that the IR scenario can be ruled out in the $x=-1/2$ model.    

Beyond the clean observables considered above, the WCs in Eqs. (\ref{WCs},\ref{WCs2}) also effect other $b\to sll$ observables. We examine the effects of these WCs using the parameter space defined in Eq. (\ref{par_space_1}). The  scalar and pseudoscalar WCs $C_{S,P}^{(')}$ are estimated to be $C_{S,P}^{(')}\sim \mathcal{O}(10^{-5})$, which are 
significantly suppressed compared to the SM WCs and can be neglected.

 For remaining six WCs  $C_{7,9,10}^{\text{NP}}$, we  note that they are identical for different lepton flavors $l=e,\mu,\tau$, implying lepton flavor university (LFU). We compare these predicted WCs with the 6D  LFU global fits at the $1\sigma$ confidence interval  \cite{Alguero:2023jeh}, as presented in Table \ref{WCs_global fit_xt12}.
\begin{table}[h]
	\protect\caption{\label{WCs_global fit_xt12} The comparison between predicted NP WCs in the case $x=-1/2$ with $1\sigma$ confidence interval in the 6D LFU global fit result \cite{Alguero:2023jeh} } 
	\begin{centering}
		\begin{tabular}{|c|c|c|c|c|c|c|}
			\hline
			WCs & $C_{9}^{\text{NP}}$  & $C_{10}^{\text{NP}}$  & $C_{9}^{'\text{NP}}$  &$C_{10}^{'\text{NP}}$ & $C_{7}^{\text{NP}}$ &$C_{7}^{'\text{NP}}$      \tabularnewline
			\hline 
			Global fits & $[-1.38, -1.03]$ &  $[-0.09, 0.22]$& $[-0.40, 0.33]$ & $[-0.25, 0.13]$ & $[-0.01, 0.02]$ & $[0, 0.03]$ 
			\tabularnewline
			\hline 
			NR scenario  & $-0.243$ &  $0.243$& $[-0.106, -0.086]$ & $[0.086, 0.106]$ & $[0.00796, 0.0241]$ & $0$ 	\tabularnewline
			\hline 
		\end{tabular}
		\par
	\end{centering}
\end{table}
The WCs $C_{9,10}^{'\text{NP}}$ lie within the $1\sigma$ intervals, while 
 $C_{7}^{(')\text{NP}}$ have ranges interfering with their corresponding global fit values. Since the WCs $C_{9,10}^{\text{NP}}$  primarily  depend on the NP scale $\La$, which  is fixed at $5 \text{TeV}$ for  $x=-1/2$, we obtain $C_{9}^{\text{NP}}\simeq -0.243$ and $C_{10}^{\text{NP}}\simeq 0.243$. However, these values exceeds the global fit upper bounds
  $C_{9}^{\text{NP-fit}}\leq -1.03$ and $C_{10}^{\text{NP-fit}}\leq 0.22$
 by approximately 76\% and 10\%, respectively.
 
  Consequently, for $x=-1/2$ with parameter space in Eq. (\ref{par_space_1}), the  model can explain the clean observables but not for other $b\to sll$ observables. 

\subsubsection{The case $x=1/2$}
In contrast to the previous scenario,  the NP scale $\La$ is now constrained to the range $\La \in [4.5, 8.5]$ TeV, while  VEV $v_1$ falls within the range $v_1\in[1,185]$ GeV, as reported  in  \cite{VanDong:2022cin}.  For the case $x=1/2$, we numerically investigate the parameter spaces that satisfy the constraints on clean observables given in  Eqs.(\ref{constraint1}),(\ref{constraint2}),(\ref{Bsmm_constraint}),(\ref{Bsg_constraint}) in both the NR and IR scenarios for mixing angles in $V_{dR}$ matrix. 

\begin{figure}[h]
	\centering
	\begin{tabular}{cc}
		\includegraphics[width=8.0cm]{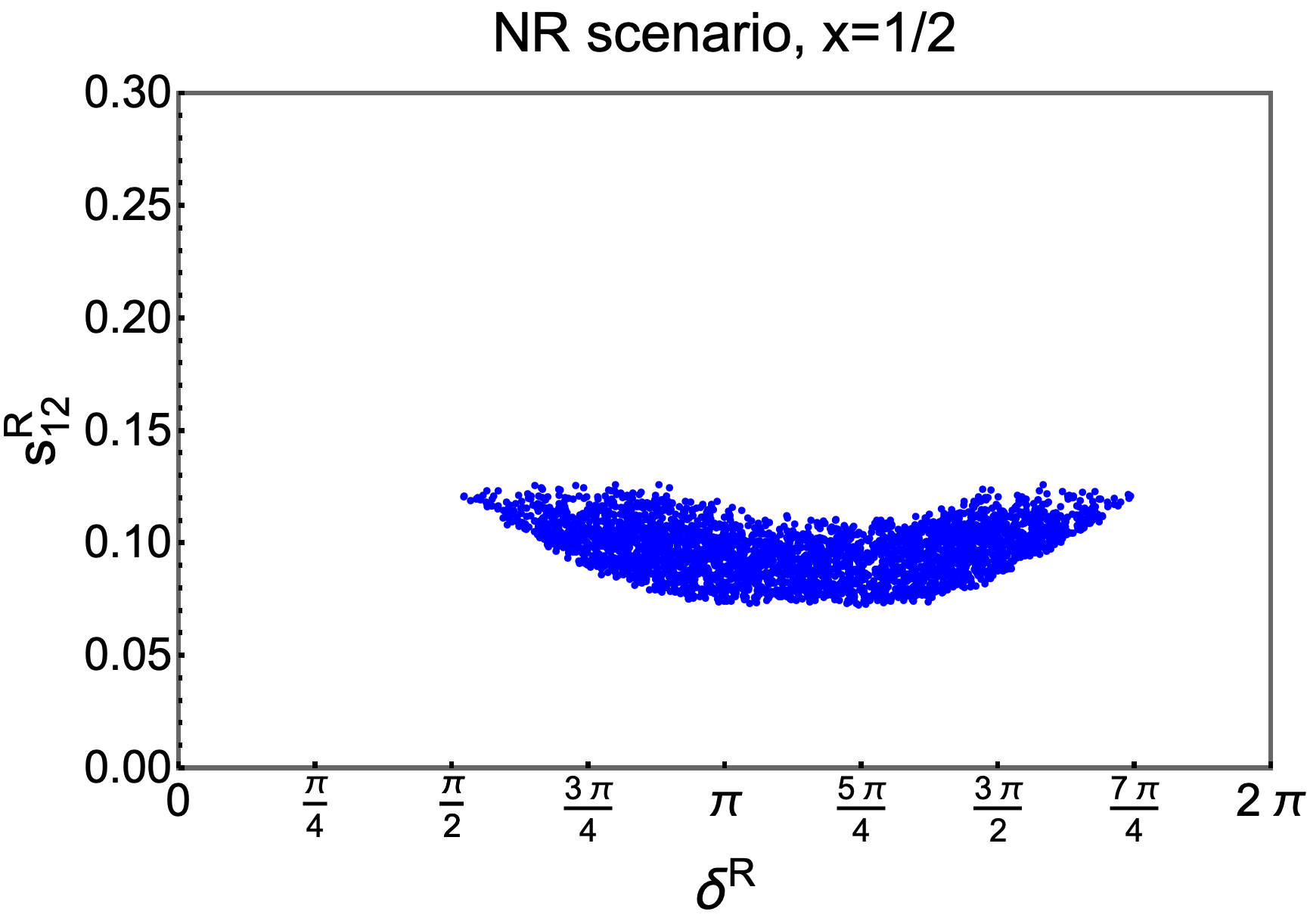}
		\includegraphics[width=8.0cm]{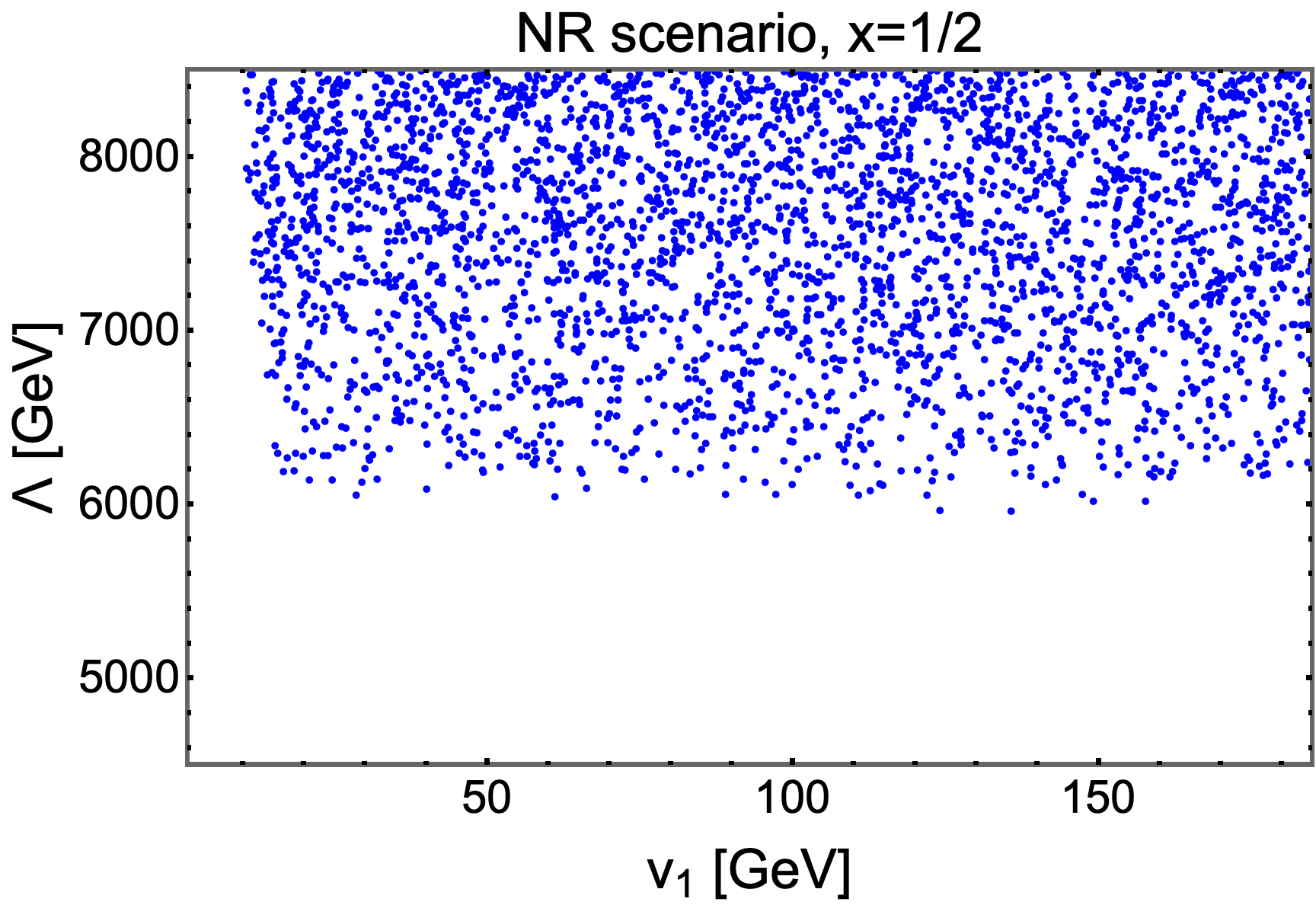}
	\end{tabular}
	\caption{ \label{xc12_NR} The left and right panels show the correlation between $s_{12}^R-\delta^R$ and $\La-v_1$, respectively. The mixing angle parameters in $V_{dR}$ matrix are assumed in the NR scenario $s_{12}^R>s_{23}^R>s_{13}^R$.}
\end{figure}
The left panel in the Fig. \ref{xc12_NR} shows the correlation between mixing angle parameter $s_{12}^R$ and CP-violating phase $\delta^R$ when relaxing the $2\sigma$ constraint on quark flavor-changing processes. We note  that the allowed region for  $s_{12}^R$ and $\delta^R$ is larger than in the $x=-1/2$ case.  Specifically, we find $s_{12}^R\in [0.074 ,0.123]$ and  $\delta^R\in[0.52\pi, 1.74\pi]$. Turning to the right panel, which presents the correlation between two VEVs $\La$ and $v_1$ while satisfying all $2\sigma$ constraint on quark flavor observables, we obtain the lower limits $v_1\geq 10.6$ GeV and $\La\geq 5964 $ GeV.

\begin{figure}[h]
	\centering
	\begin{tabular}{cc}
		\includegraphics[width=8.0cm]{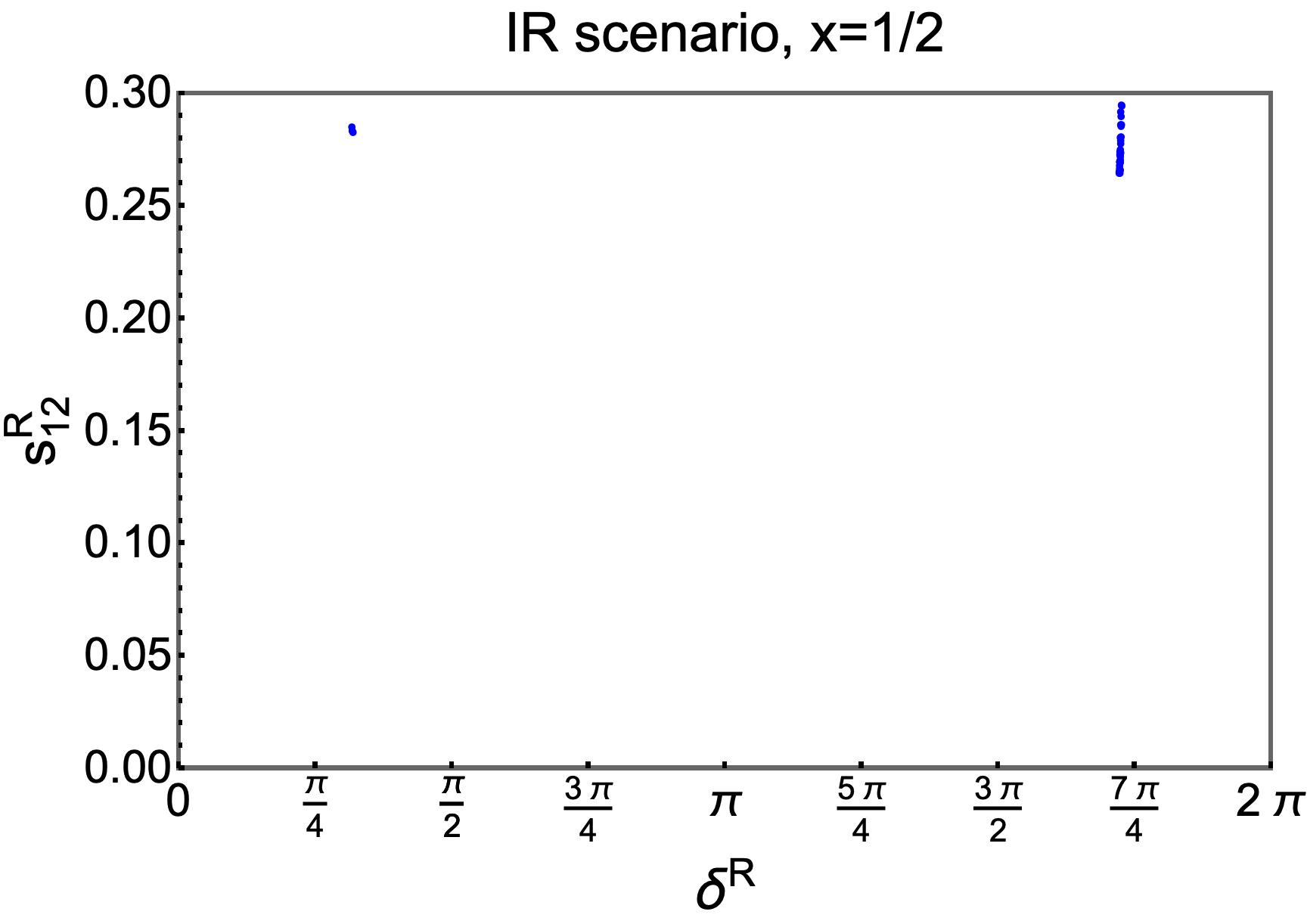}
		\includegraphics[width=8.0cm]{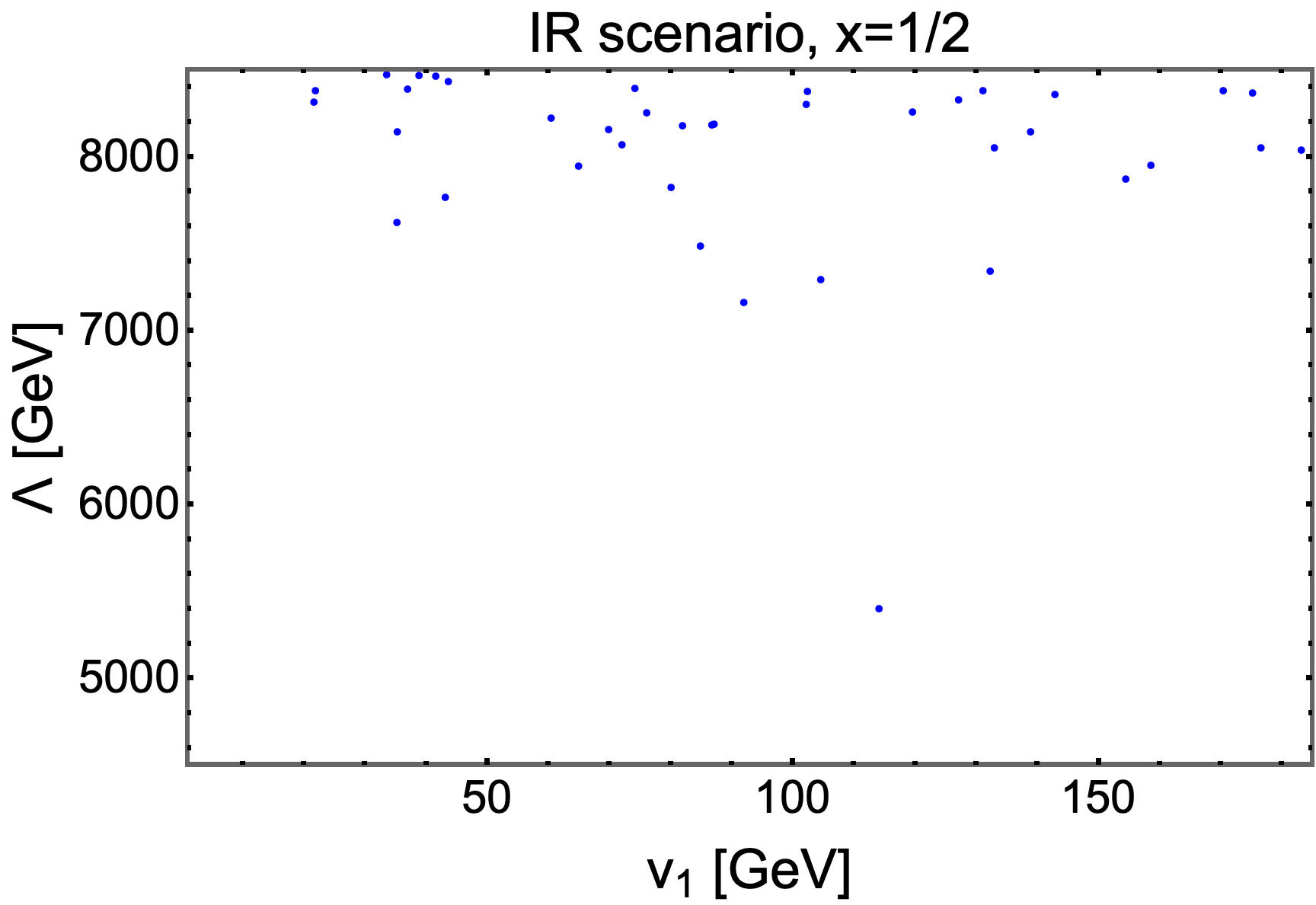}
	\end{tabular}
	\caption{ \label{xc12_IR} The left and right panels show the correlation between $s_{12}^R-\delta^R$ and $\La-v_1$, respectively. The mixing angle parameters in $V_{dR}$ matrix are assumed in the IR scenario $s_{12}^R>s_{13}^R>s_{23}^R$.}
\end{figure}

For the IR scenario with $s_{12}^R>s_{13}^R>s_{23}^R$, we conducted a numerical study similar to the NR case and obtain the results shown in Figs.(\ref{xc12_IR}).

Figs. \ref{xc12_IR} illustrates the correlation between $s_{12}^R-\delta^R$, and $\La-v_1$ in the $x=1/2$ case with IR condition on the mixing angle parameters in $V_{dR}$ matrix. Compared to NR case, this scenario yields significantly fewer points that satisfy the constraints of flavor observations. Especially, the left panel reveals remarkably narrow regimes for $\delta^R$ around $\delta^R\sim 0.32\pi$ and
$\delta^R\sim 1.72\pi$, while the mixing angle $s_{12}^R\sim 0.28$ and $s_{12}^R\sim [0.265,0.295]$, respectively. These obtained values for $s_{12}^R$ are slightly larger than the central value of the relative CKM mixing angle $s_{12}^{\text{CKM}}\sim 0.22548$ \cite{Charles:2015gya}. In  the right panel, the points are primarily concentrated in the region where $\La\geq 7200$ GeV and $v_1\geq 22$ GeV. These lower bounds for the  VEVs are stringent  than in the NR case. 

Using the free parameters obtained from the NR and IR scenarios, we can estimate the impact of the non-standard model WCs  $C_{7,9,10, S,P}^{(\prime)\text{NP}}$ on other $b\to sll$ observables. Numerical analysis reveal that  scalar and pseudoscalar WCs $|C_{S,P}^{(\prime)}|$, with magnitudes of$ \mathcal{O}(10^{-4}-10^{-5})$,  are significantly suppressed compared to the SM WCs, $C_{7,9,10}^{\text{SM}}\sim \mathcal{O}(10^{-1})$.   Consequently, the effects of new scalars can be safely neglected. This  leaves six NP WCs $C_{7,9,10}^{(\prime)\text{NP}}$ contributing to $b\to sll$ observables. This result aligns with $x=-1/2$ case. The predicted results and the 6D LFU global fit \cite{Alguero:2023jeh} are shown in Table \ref{WCs_global fit_xc12}.

\begin{table}[h]
	\protect\caption{\label{WCs_global fit_xc12} The comparison between predicted NP WCs in the case $x=1/2$ with $1\sigma$ confidence interval in the 6D LFU global fit results \cite{Alguero:2023jeh}}
	\begin{centering}
		\begin{tabular}{|c|c|c|c|c|c|c|}
			\hline
			WCs & $C_{9}^{\text{NP}}$  & $C_{10}^{\text{NP}}$  & $C_{9}^{'\text{NP}}$  &$C_{10}^{'\text{NP}}$ & $C_{7}^{\text{NP}}$ &$C_{7}^{'\text{NP}}$      \tabularnewline
			\hline 
			Global fits & $[-1.38, -1.03]$ &  $[-0.09, 0.22]$& $[-0.40, 0.33]$ & $[-0.25, 0.13]$ & $[-0.01, 0.02]$ & $[0, 0.03]$ 
			\tabularnewline
			\hline 
			NR scenario  & $[-1.19, -0.59]$ &  $[-0.17, -0.08]$& $[-0.55, -0.19]$ & $[-0.08, -0.03]$ & $[1.1\times 10^{-4}, 2.41\times 10^{-2}]$ & $0 $ 	\tabularnewline
			\hline 
			IR scenario  & $[-1.45, -0.59]$ &  $[-0.21, -0.08]$& $[-0.99, -0.40]$ & $[-0.14, -0.06]$ & $[1.26\times 10^{-4}, 9.79\times 10^{-3}]$ & $0$ 	\tabularnewline
			\hline 
		\end{tabular}
		\par
	\end{centering}

\end{table}

As shown in Table \ref{WCs_global fit_xc12},  for NR scenario, the predicted interval for $C_{10}^{'\text{NP}}$ falls within the $1\sigma$ global fit range for $C_{10}^{'\text{NP-fit}}$. However, the remaining $C_{9}^{(')\text{NP}}$, $C_{10}^{\text{NP}}$ and $C_{7}^{(')\text{NP}}$  exhibit interference with their corresponding global fit values . In contrast, for IR scenario,  the predicted interval for $C_{9}^{'\text{NP}}$ does not overlap with the $1\sigma$ global  fit for $C_{9}^{'\text{NP-fit}}$. Specifically, the maximum  value of $C_{9}^{'\text{NP}}\leq -0.4$, while the global fit implies $C_{9}^{'\text{NP-fit}}\geq -0.4$. Consequently, the model with  $x=1/2$ and NR  mixing angles in $V_{dR}$ matrix can successfully explain both clean observables and other $b\to sll$ observables.  However, the IR scenario fails to account for $b\to sll$ observables, despite explaining the clean observables.
\subsection{New physics scale is limited by CMS measurement of the $W$ gauge boson mass  }
Given the recent CMS measurement of $W$ gauge  boson mass,  which is now consistent with SM prediction \cite{CMSMW2024} \cite{Bona:2018}, the model suggests that  all four cases of $x$ charge parameter $x={\pm 1/6,\pm 1/2}$ can accommodate  this new CMS measurement and the global fit of the $\rho$ parameter. In this section, we will revisit the flavor observables inspired by constraints on the NP scale  arising from  the CMS measurement of the $W$ gauge boson mass .
\begin{figure}[h]
	\centering
	\begin{tabular}{cc}
		\includegraphics[width=8.0cm]{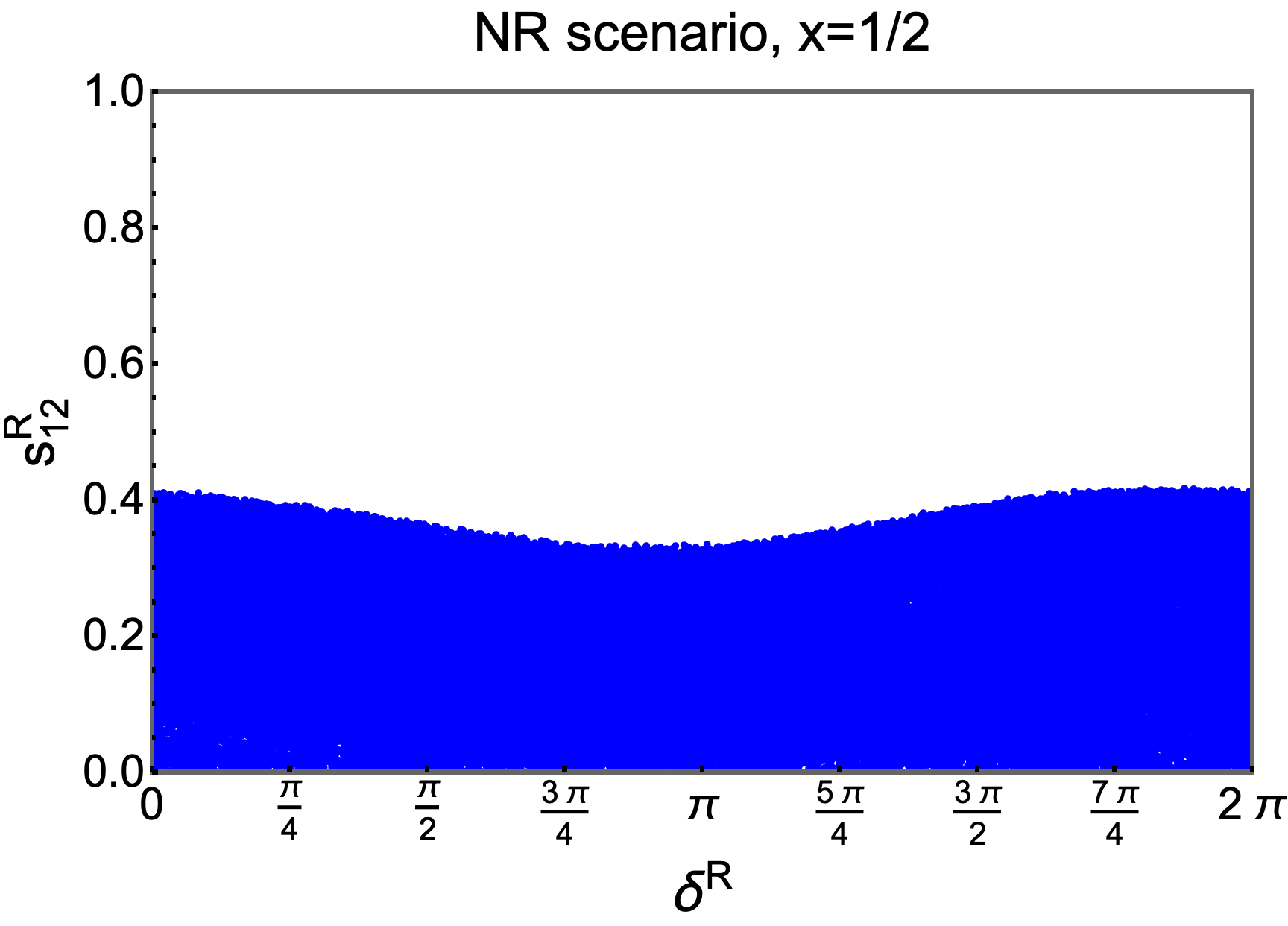}
		\includegraphics[width=8.0cm]{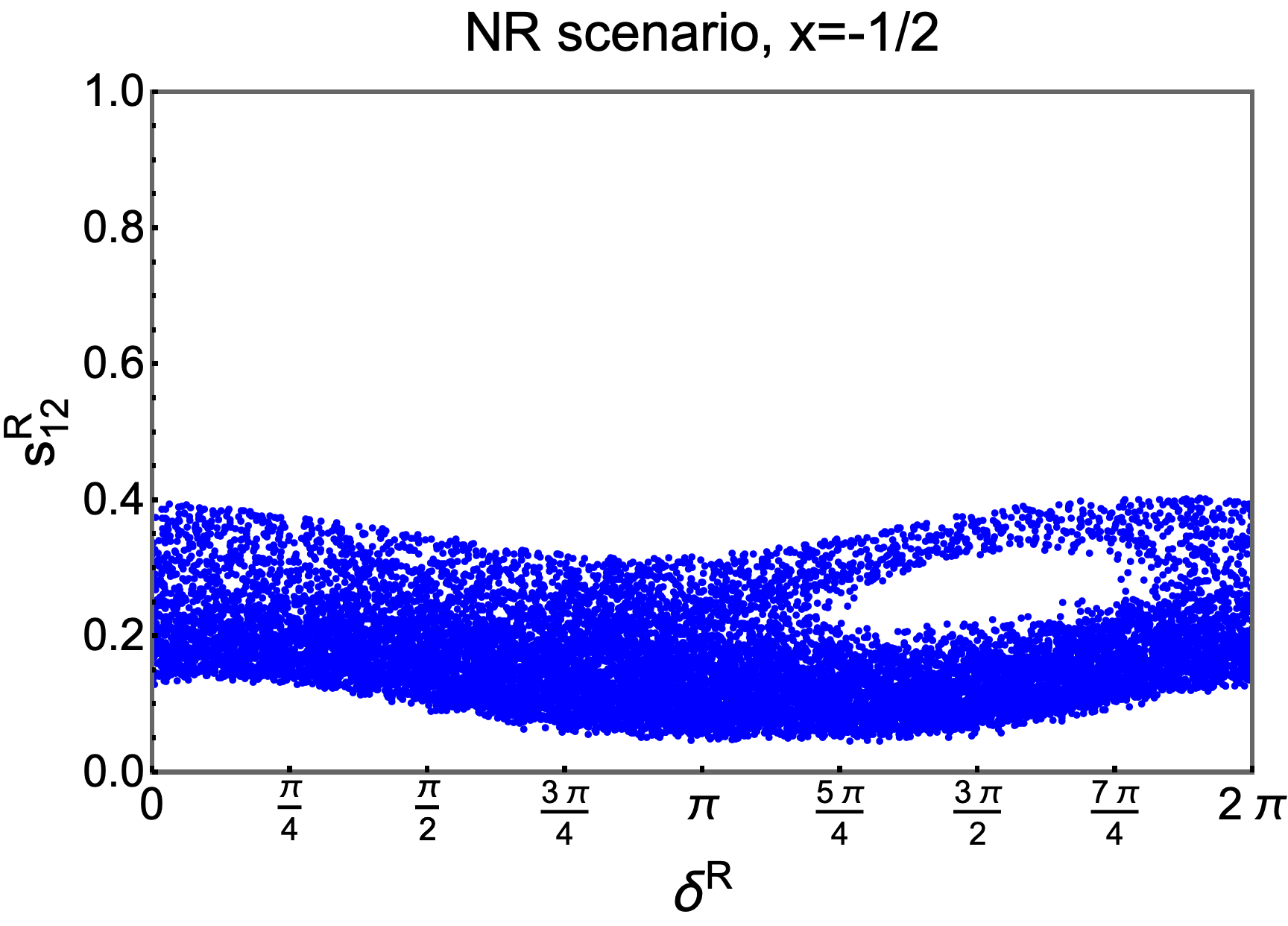}\\ 
			\includegraphics[width=8.0cm]{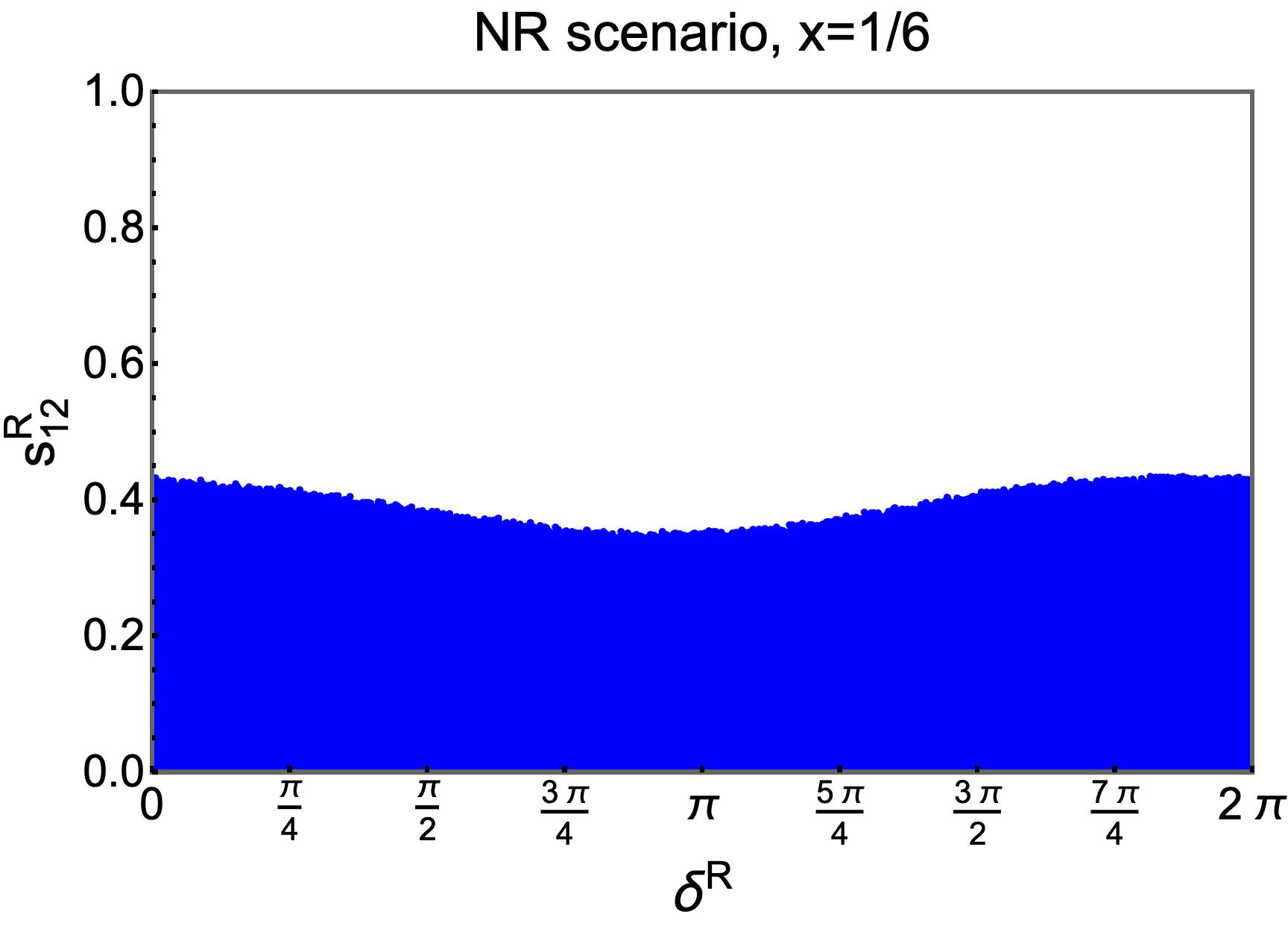}
		\includegraphics[width=8.0cm]{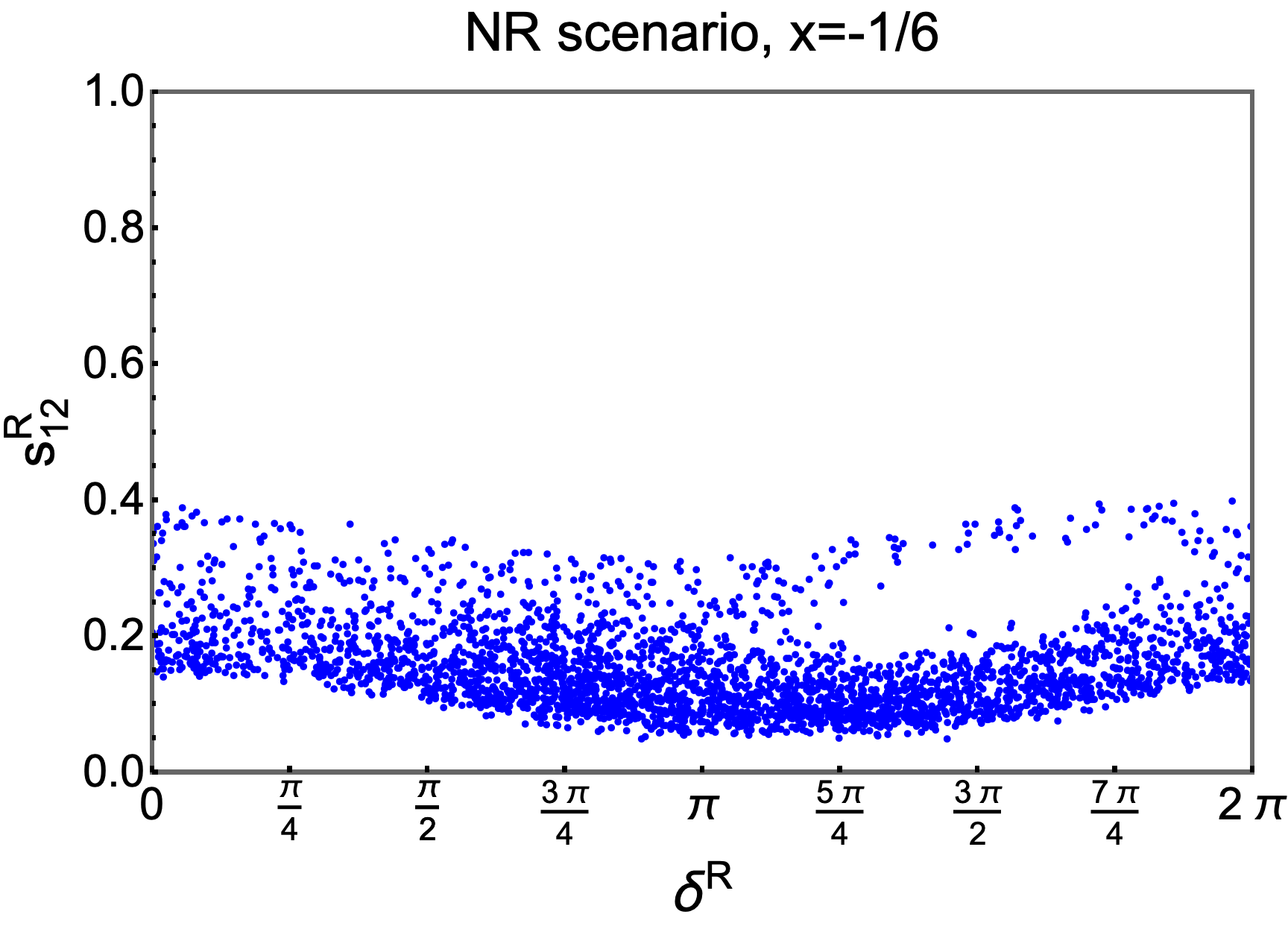} 
	\end{tabular}
	\caption{ \label{MWCMS_NR_1} Top and bottom panels show the correlation between $s_{12}^R-\delta^R$ for $x=\pm1/2$ and $x=\pm1/6$, respectively. The mixing angle parameters in $V_{dR}$ matrix are assumed in the NR scenario $s_{12}^R>s_{13}^R>s_{23}^R$.}
\end{figure}
\begin{figure}[h]
	\centering
	\begin{tabular}{cc}
		\includegraphics[width=8.0cm]{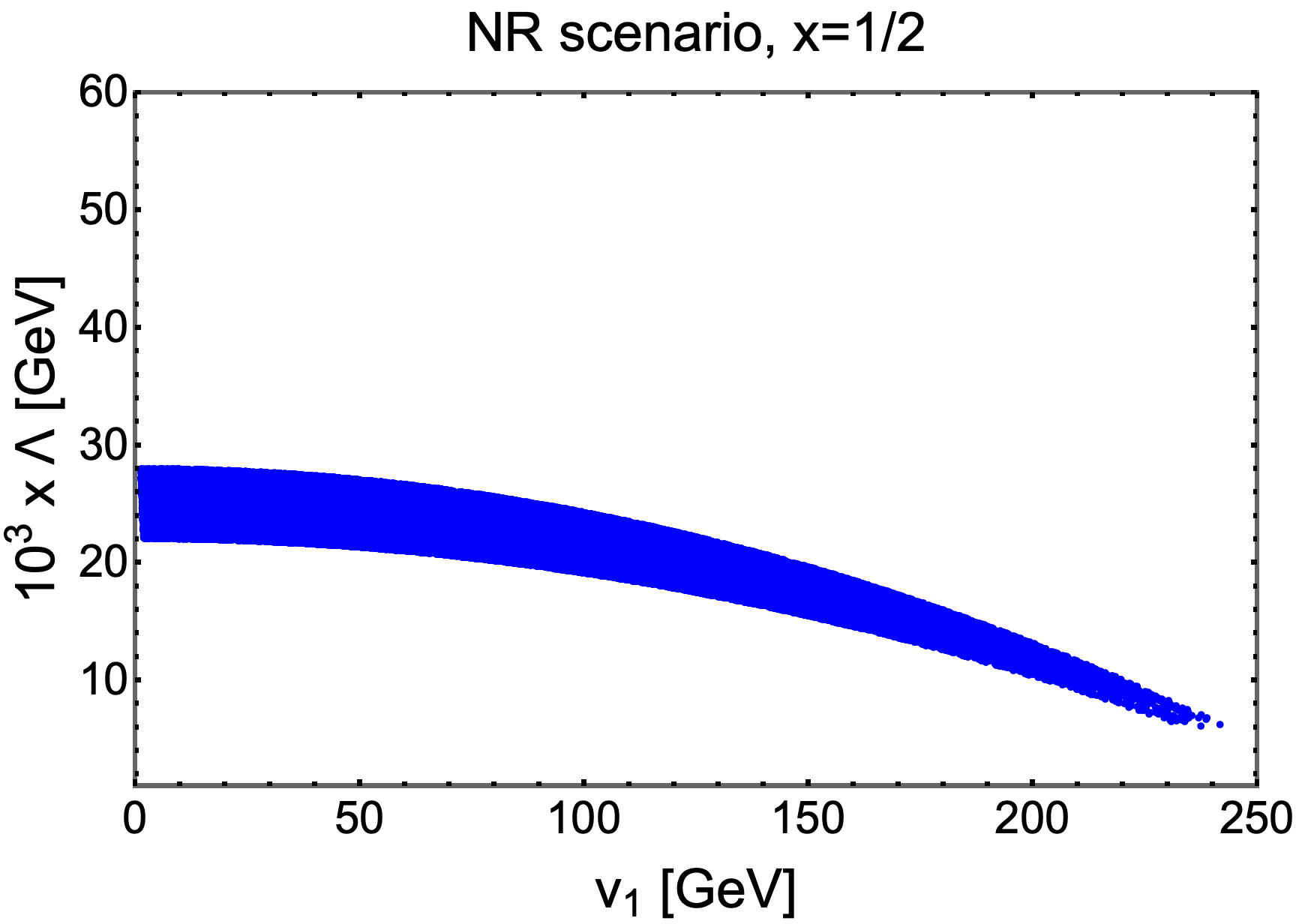}
		\includegraphics[width=8.0cm]{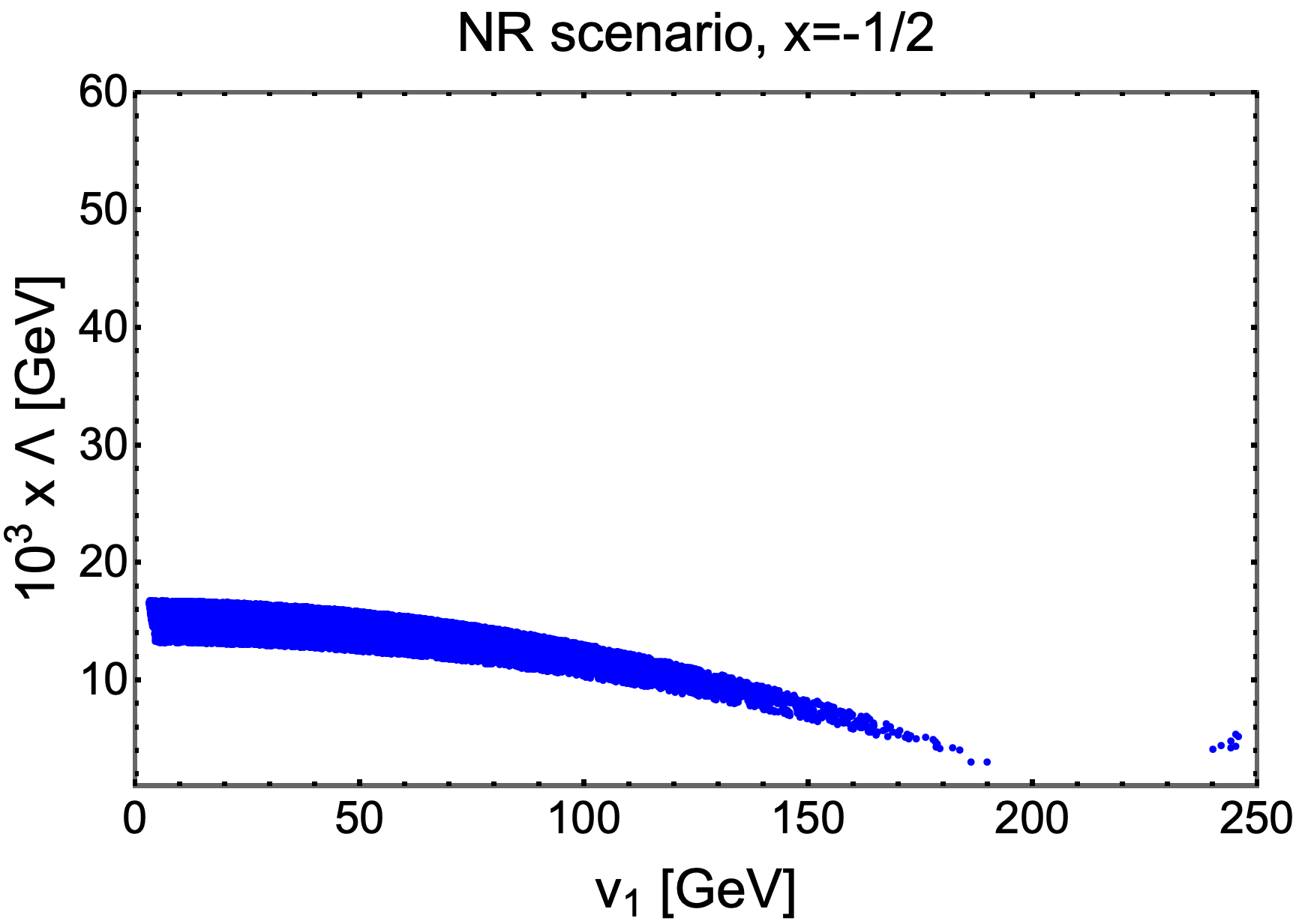}\\ 
		\includegraphics[width=8.0cm]{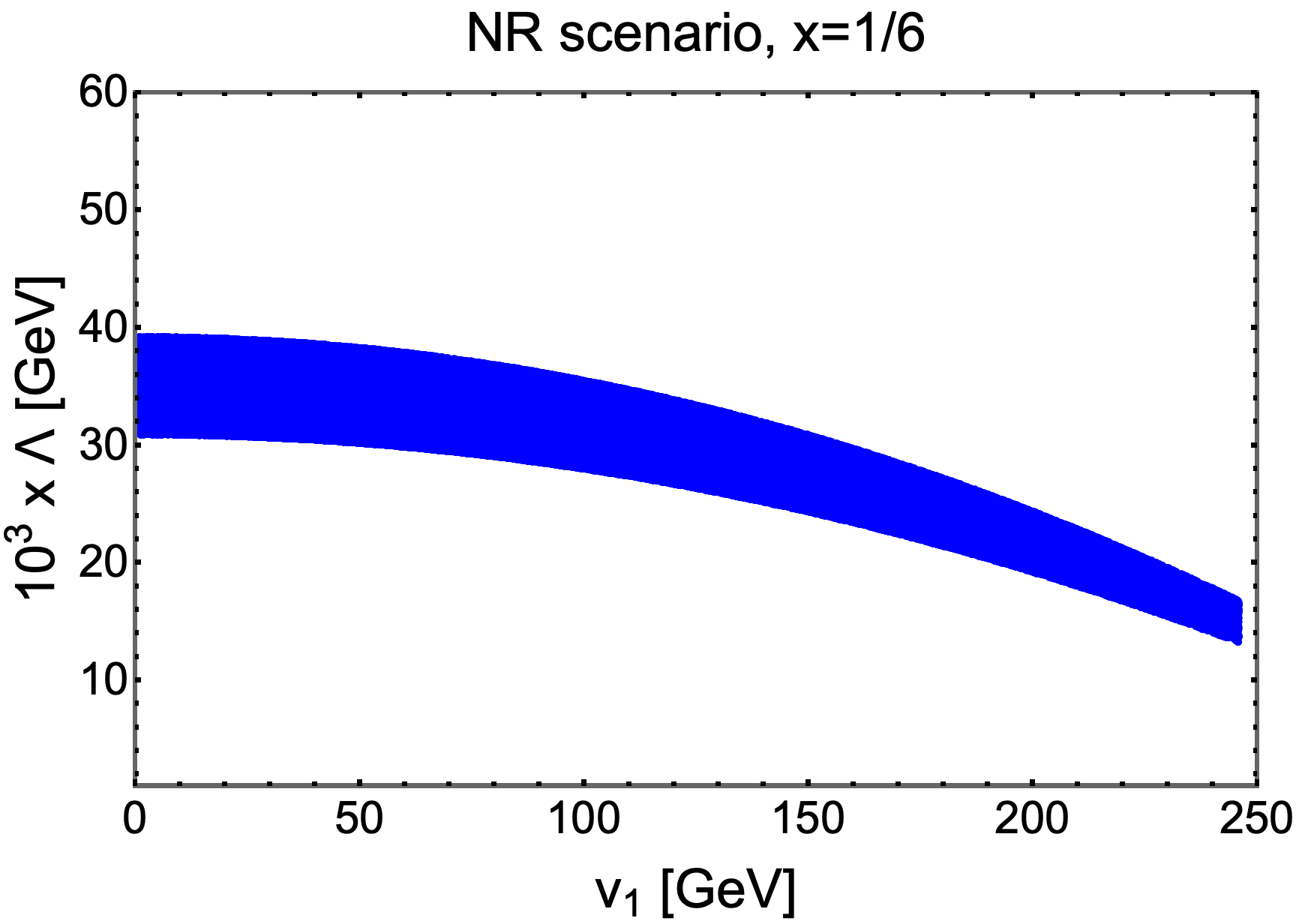}
		\includegraphics[width=8.0cm]{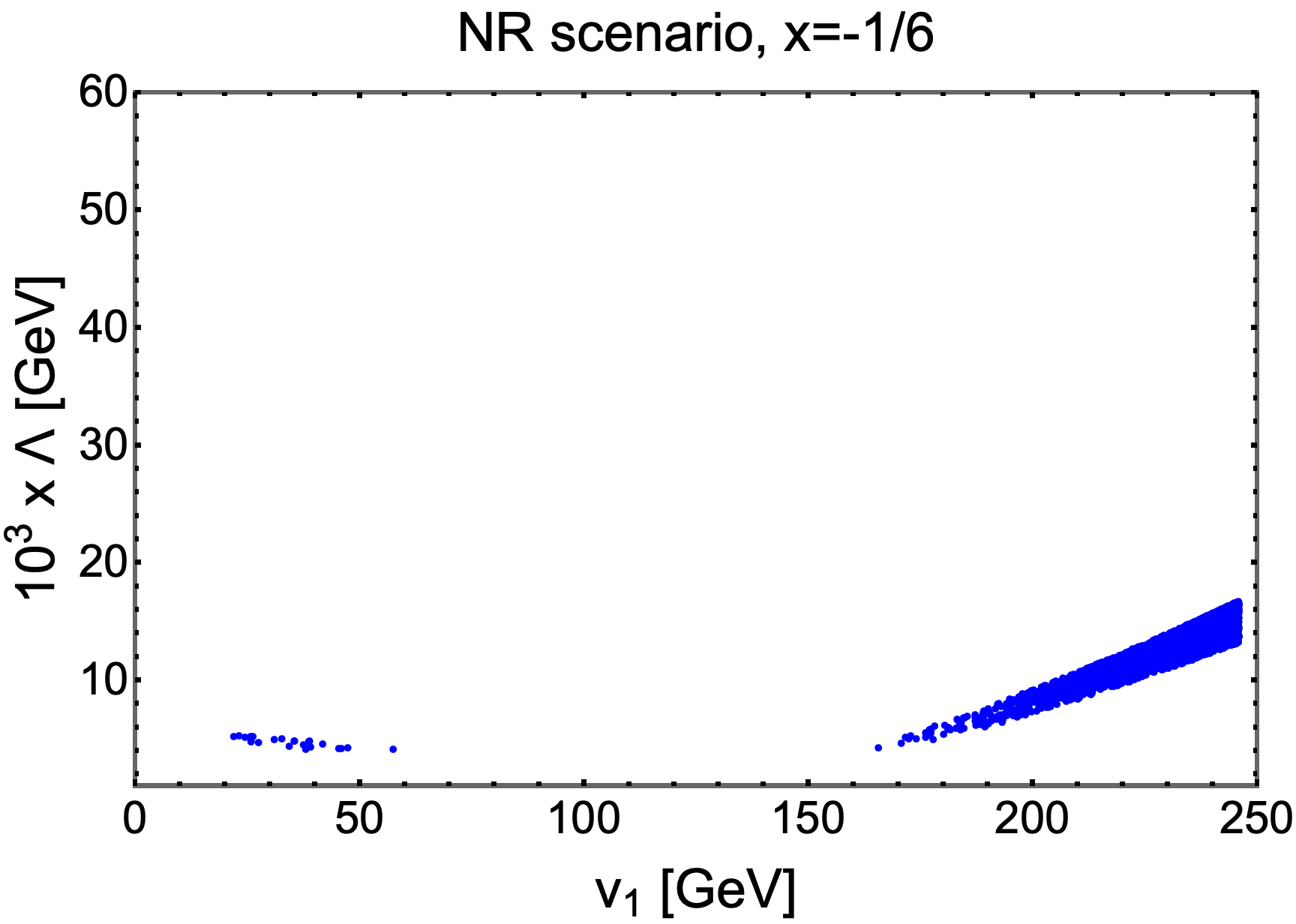} 
	\end{tabular}
	\caption{ \label{MWCMS_NR_2} Top and bottom panels show the correlation between $v_1-\La$ for $x=\pm 1/2$ and $x=\pm1/6$, respectively. The mixing angle parameters in $V_{dR}$ matrix are assumed in the NR scenario $s_{12}^R>s_{13}^R>s_{23}^R$.}
\end{figure}
In the panels corresponding four cases  $x=\pm 1/2$ and $x=\pm 1/6$ of Figs. \ref{MWCMS_NR_1}, we present the correlations between the mixing angle $s_{12}^{R}$ and CP violating phase $\delta^R$ satisfying constraints on quark flavor observables, with the mixing angles of $V_{d_R}$ set to  NR values.  

All four choices of $x$  allow for regions thats satisfy given flavor constraints. Notably, the obtained ranges for $s_{12}^R$ and $\delta^R$ are larger than in previous studies, For instance, we find  $\delta^R\in[0,2\pi]$ and $s_{12}^R\in[0, 0.4]$ for cases $x=1/2,1/6$ and $s_{12}^R\in[0.05,0.4]$ for $x=-1/2,-1/6$. This behavior can be attributed to the new physics scale $\La$  being significantly larger than in previous studies based on CDF II results. As a result, the NP contributions are more suppressed, allowing for a wider range of acceptable parameter values.

The panels in Figs .\ref{MWCMS_NR_2} illustrate  the relationship between two VEVs $v_1$ and $\La$ satisfying flavor constraints. Comparing these panels with these in Figs .\ref{MWCMS}, we observe that the behavior for 
 $x=1/2,1/6$ is quite consistent, whereas the panels for  $x=-1/2,-1/6$ exhibit two distinct regions.  Specifically, there are narrow regimes with $\La\sim 2-4$ TeV and  $v_1\in [240,246]$ GeV, as well as $\La\sim 5$ TeV and $v_1\in [20,50]$ GeV for $x=-1/2$ and $x=-1/6$, respectively.  Notably, the panels for $x=-1/2$ and $x=-1/6$ in Figs. \ref{MWCMS_NR_1} exhibit excluded regions $\delta^R\in [5\pi/4, 7\pi/4]$ and $s_{12}^R\in [0.2,0.3]$. These regions arise due to the discontinuous nature of the allowed 
$v_1-\La$ parameter space in the corresponding panels of Figs .\ref{MWCMS_NR_2}. Specifically, there are gaps in the allowed region for $v_1$ around $v_1\in[190, 240]$ GeV and $v_1\in[60, 160]$ GeV, which prevent the model from satisfying the flavor constraints. 

\begin{figure}[h]
	\centering
	\begin{tabular}{cc}
		\includegraphics[width=8.0cm]{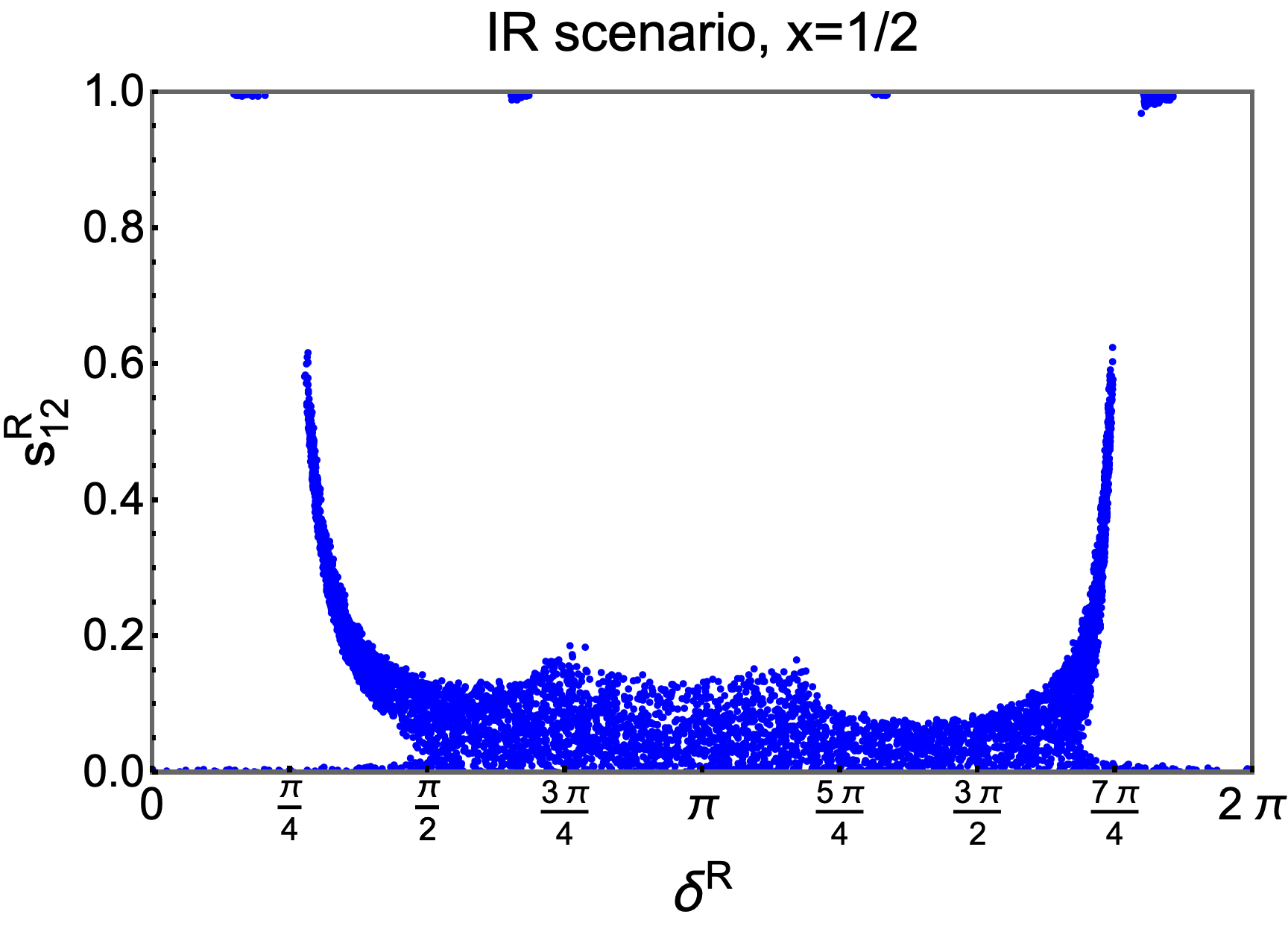}
		\includegraphics[width=8.0cm]{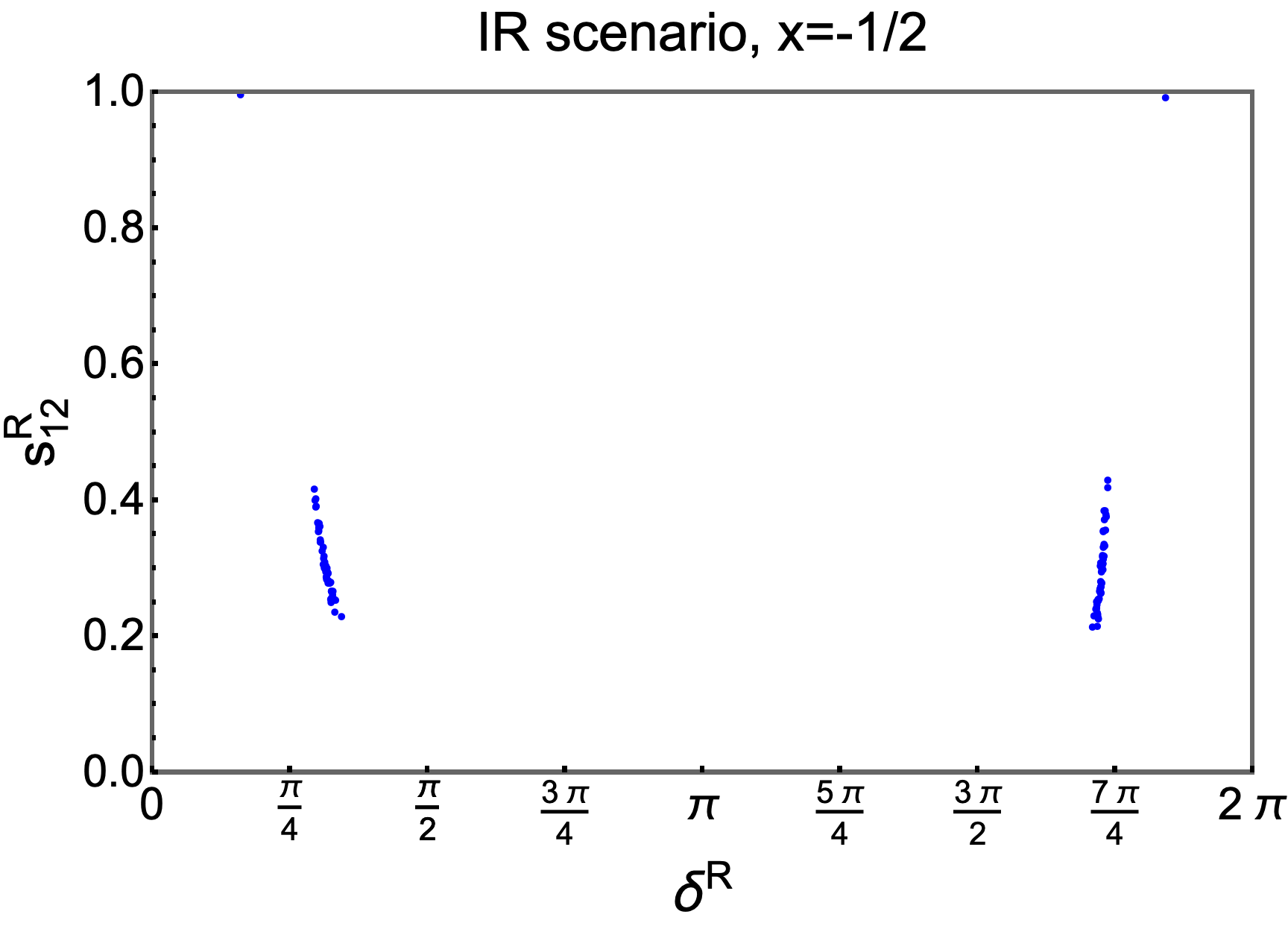}\\ 
		\includegraphics[width=8.0cm]{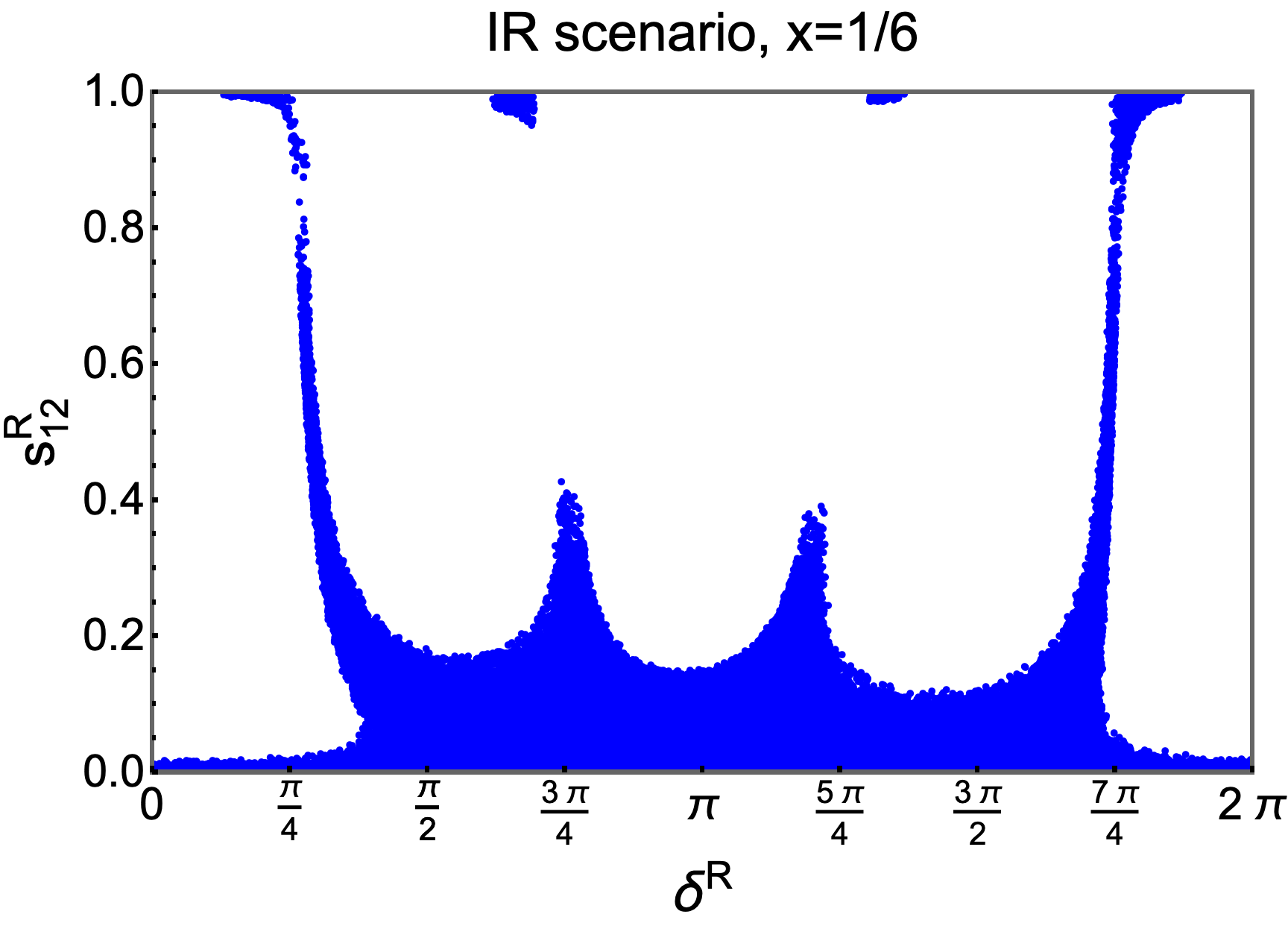}
		\includegraphics[width=8.0cm]{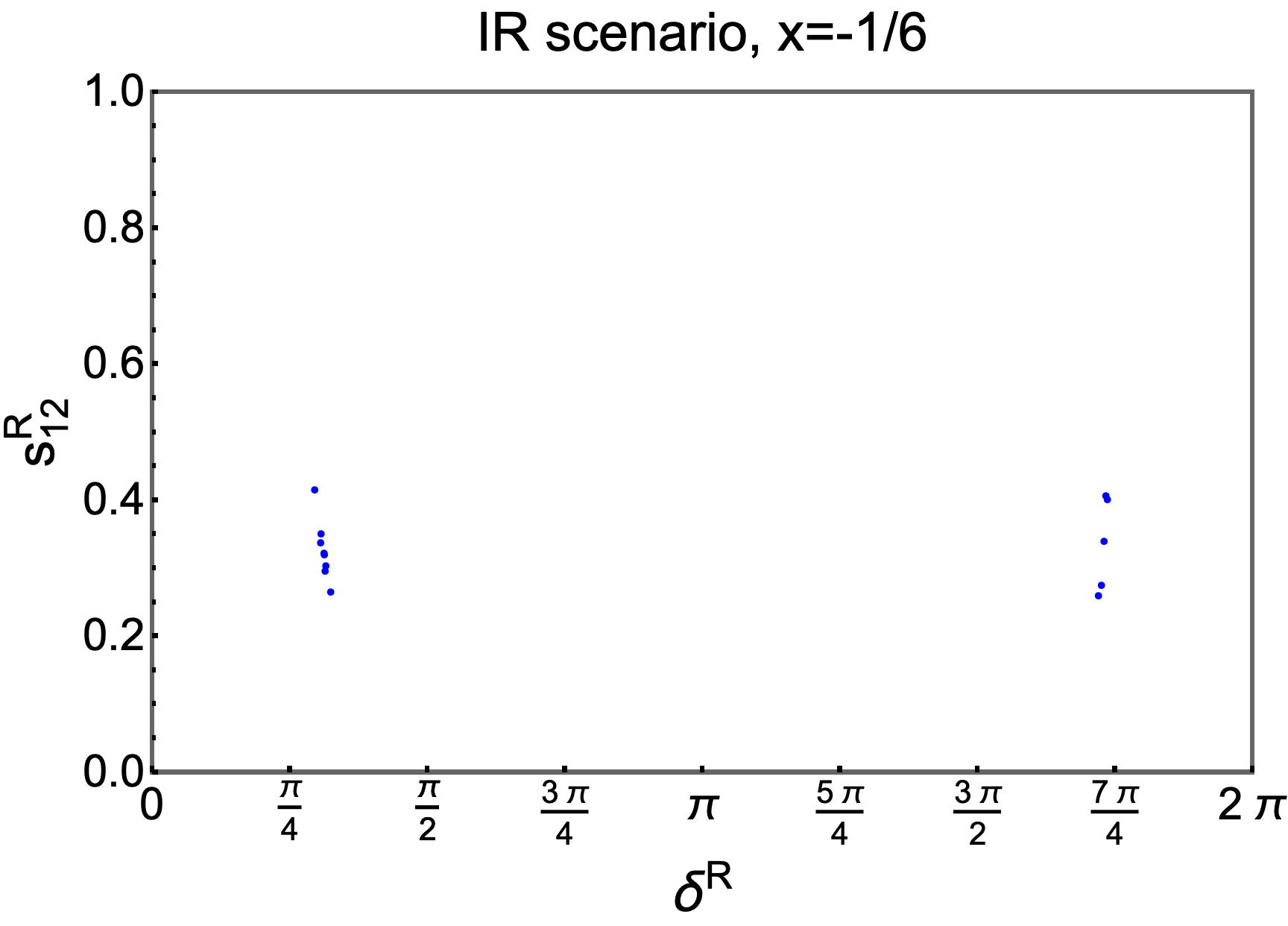} 
	\end{tabular}
	\caption{ \label{MWCMS_IR_1} Top and bottom panels show the correlation between $s_{12}^R-\delta^R$ for $x=\pm1/2$ and $x=\pm1/6$, respectively. The mixing angle parameters in $V_{dR}$ matrix are assumed in the IR scenario $s_{12}^R>s_{13}^R>s_{23}^R$.}
\end{figure}

\begin{figure}[h]
	\centering
	\begin{tabular}{cc}
		\includegraphics[width=8.0cm]{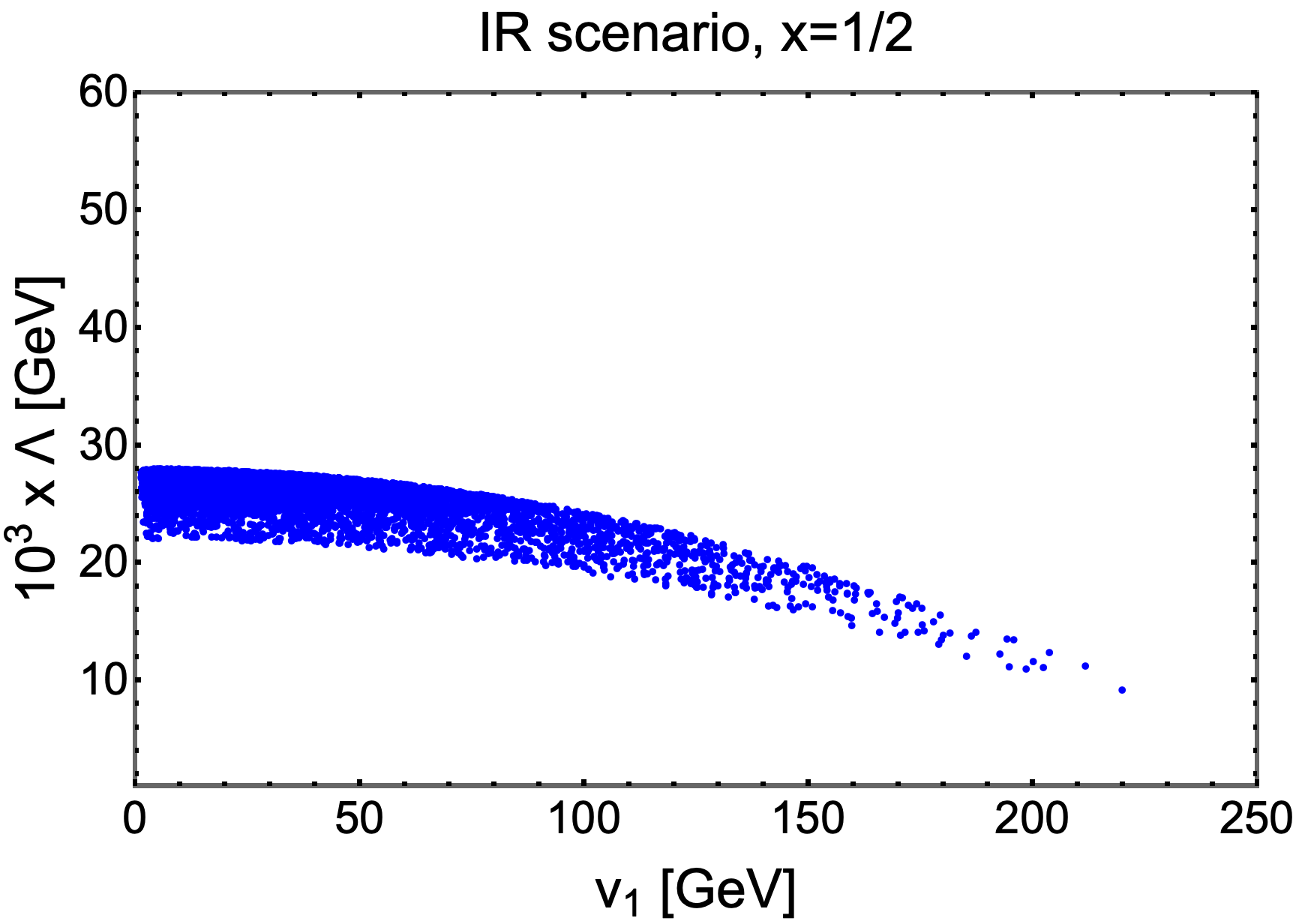}
		\includegraphics[width=8.0cm]{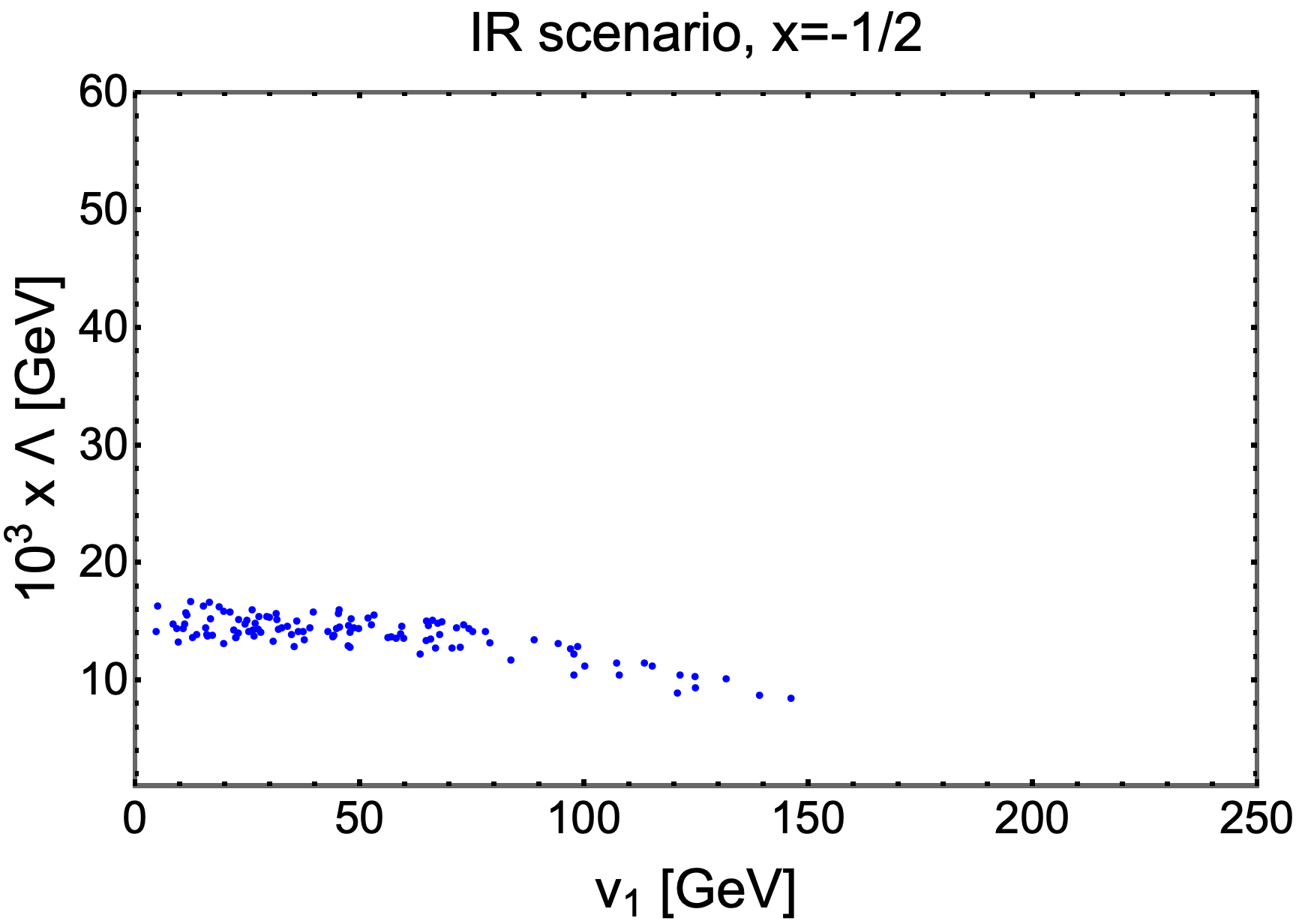}\\ 
		\includegraphics[width=8.0cm]{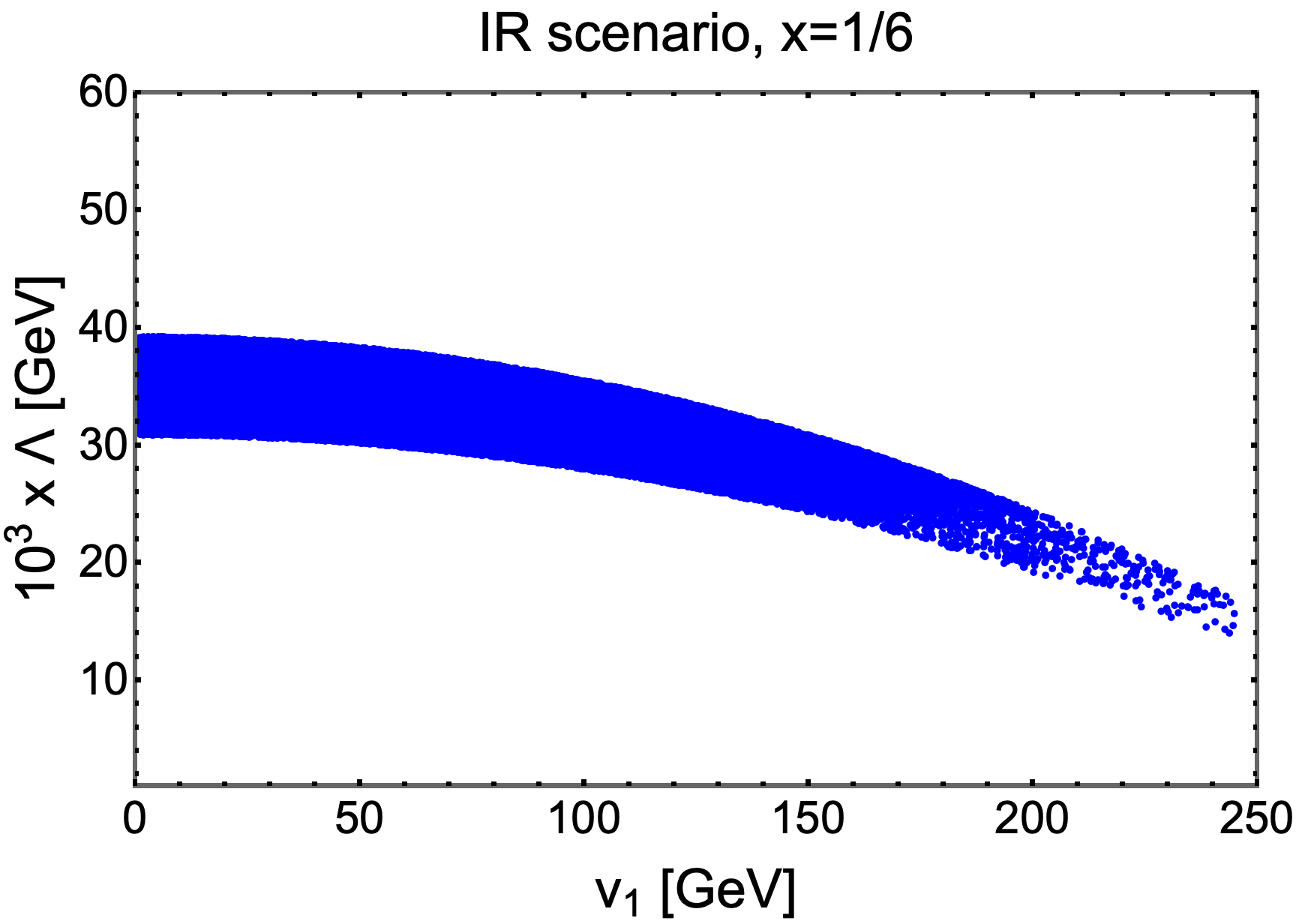}
		\includegraphics[width=8.0cm]{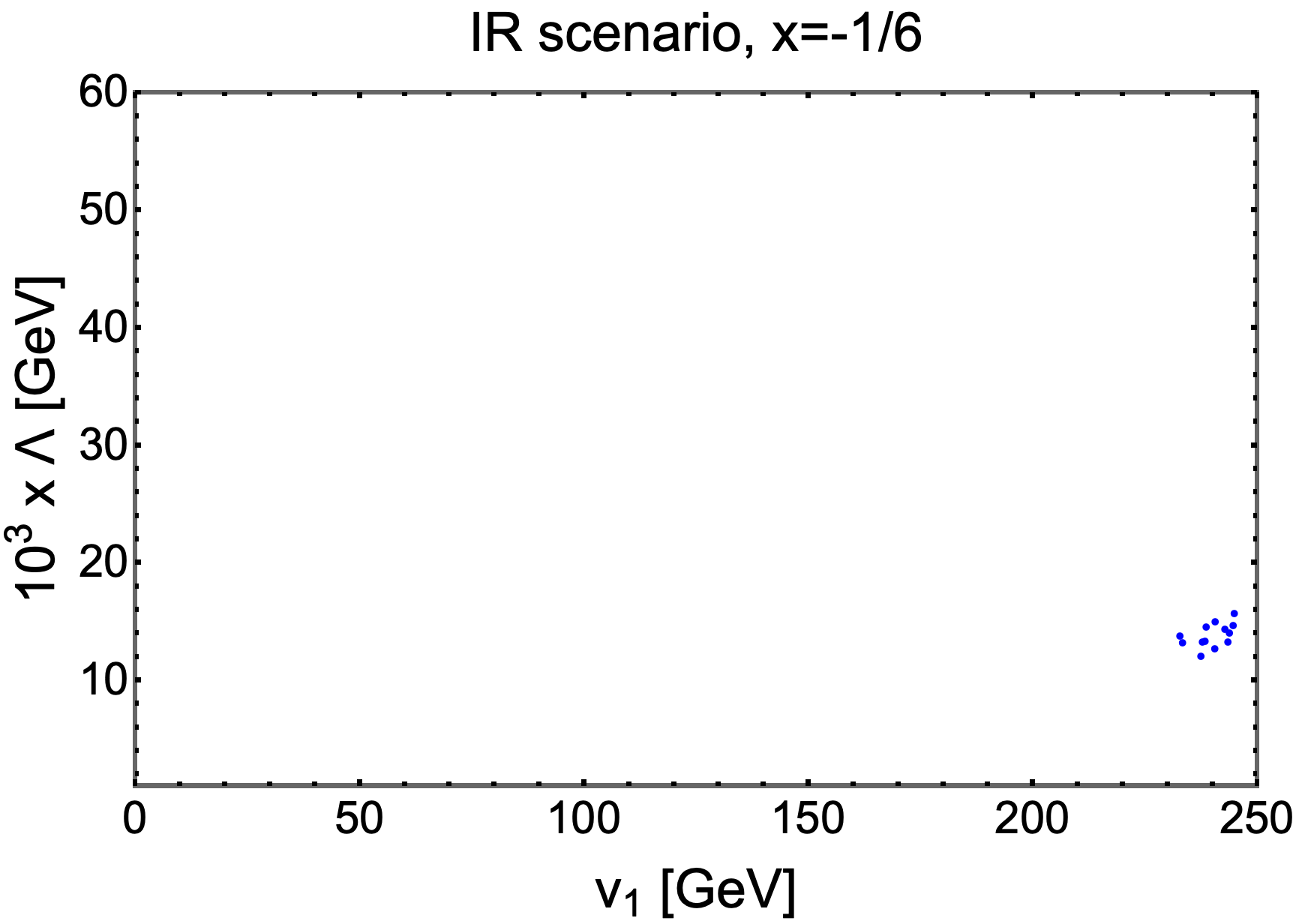} 
	\end{tabular}
	\caption{ \label{MWCMS_IR_2} Top and bottom panels show the correlation between $v_1-\La$ for $x=\pm 1/2$ and $x=\pm1/6$, respectively. The mixing angle parameters in $V_{dR}$ matrix are assumed in the NR scenario $s_{12}^R>s_{13}^R>s_{23}^R$.}
\end{figure}

Next, we focus on the IR scenario, as depicted in Fig .\ref{MWCMS_IR_1} and \ref{MWCMS_IR_2}. In Fig .\ref{MWCMS_IR_1}, we observe a stronger correlation between the mixing angle $s_{12}^R$ and the CP-violating phase $\delta^R$ compared to the NR scenario. For  $x=-1/2$ and $x=-1/6$, the allowed region for $s_{12}^R$ is significantly constrained to  $s_{12}^R\sim [0.2,0.4]$ with $\delta^R\sim 3\pi/4$ or $\delta^R\sim 7\pi/4$; In contrast,  for $x=1/2$ and $1/6$, the allowed regions for $\delta^R$ is primarily limited to $\delta^R \in [\pi/4,7\pi/4]$, while  $s_{12}^R$ can attain maximum values of $s_{12}^R\sim 1$ at specific values  of $\delta^R\sim \left\{\pi/4,3\pi/4,11\pi/8,7\pi/4\right\}$.  In Fig .\ref{MWCMS_IR_2}, the constraints on the VEVs $v_1$ and $\La$ are more stringent in the IR scenario compared to the NR scenario, especially for  $x=-1/2$ and $-1.6$. For example,  we find limits such as $v_1\in[0,150]$ GeV, $\La \in [10,16]$ TeV and $v_1\in [230,246]$ GeV, $\La \in [12,16]$ TeV.

Finally, we examine the impact of the obtained parameter spaces on the non-standard model WCs. 
As shown in Table \ref{WCs_global fit_NR_CMS},  for NR scenario, the choices $x=-1/2$ and $x=\pm 1/6$ are ruled out due to predicted ranges of $C_9^{\text{NP}}$ that significantly exceed the  $1\sigma$ global fit range for $C_9^{\text{global-fit}}$. In contrast, the  $x=1/2$ case can potentially explain both clean and other $b\to sl^+l^-$ obsevables. Turning to IR situation, as presented in Table \ref{WCs_global fit_IR_CMS}, all four cases $x=\pm 1/2,\pm 1/6$ conflict with global fit for $C_9^{\text{NP}}$. If we revisit flavor constraints inspired by current CMS result of $m_W$, the model with  $x=1/2$ and NR mixing angles of $V_{d_R}$ can successfully explain  both clean and other $b \to sl^+l^-$ observables, while the IR scenario is excluded for all  values of $x$. This finding aligns with previous studies, indicating that impact  of CMS measurement on $m_W$ does not significantly alter the overall conclusions related to flavor changing procsesses.

\begin{table}[h]
	\protect\caption{\label{WCs_global fit_NR_CMS} The comparison between predicted NP WCs in the case $x=\pm1/2,\pm 1/6$ with $1\sigma$ confidence interval in the 6D LFU global fit results \cite{Alguero:2023jeh} }
\begin{centering}
		\begin{tabular}{|c|c|c|c|c|c|}
			\hline
		WCs & Global fits & \multicolumn{4}{c|}{\text{NR scenario}}\\
		\cline{3-6}
		&&$x=1/2$&$x=-1/2$&$x=1/6$&$x=-1/6$\\
		\hline
			 $C_{9}^{\text{NP}}$  & $[-1.38, -1.03]$& $[-1.12,-0.054]$ & $[-0.64,-0.021]$ & $[-0.448,-0.051]$ & $[0.109,1.737]$ \tabularnewline\hline   
			 $C_{10}^{\text{NP}}$  & $[-0.09, 0.22]$ &$[-0.16,-0.0073]$ & $[0.021,0.63]$&  $[-0.103,-0.012]$ & $[0.065,1.04]$\tabularnewline\hline   
			 $C_{9}^{'\text{NP}}$  &  $[-0.40, 0.33]$& $[-0.49,0]$ & $[-0.24,-0.005]$ & $[-0.574,0]$ & $[0.024,0.675]$\tabularnewline\hline  
			 $C_{10}^{'\text{NP}}$ & $[-0.25, 0.13]$ & $[-0.071,0]$ & $[0.005,0.24]$ & $[-0.133,0]$ & $[0.015,0.404]$\tabularnewline\hline  
			 $C_{7}^{\text{NP}}$ &$[-0.01, 0.02]$& $[0, 0.024]$ & $[0,0.024]$ & $[0,0.024]$ &$[0,0.022]$\tabularnewline\hline   
			 $C_{7}^{'\text{NP}}$    & $[0, 0.03]$  & $0 $ &0 & 0&0  \tabularnewline \hline 

			\hline 
		\end{tabular}
		\par
	\end{centering}

\end{table}

\begin{table}[h]
	\protect\caption{\label{WCs_global fit_IR_CMS} The comparison between predicted NP WCs in the case $x=\pm1/2,\pm 1/6$ with $1\sigma$ confidence interval in the 6D LFU global fit results  \cite{Alguero:2023jeh}}
	\begin{centering}
		\begin{tabular}{|c|c|c|c|c|c|}
			\hline
		WCs & Global fits & \multicolumn{4}{c|}{\text{IR scenario}}\\
		\cline{3-6}
		&&$x=1/2$&$x=-1/2$&$x=1/6$&$x=-1/6$\\
		\hline
			$C_{9}^{\text{NP}}$  & $[-1.38, -1.03]$& $[-0.501,-0.054]$ & $[-0.084,-0.022]$ & $[-0.4,-0.051]$ & $[0.122,0.208]$ \tabularnewline\hline   
			$C_{10}^{\text{NP}}$  & $[-0.09, 0.22]$ &$[-0.072,-0.008]$ & $[0.022,0.84]$&  $[-0.092,-0.012]$ & $[0.073,1.25]$\tabularnewline\hline   
			$C_{9}^{'\text{NP}}$  &  $[-0.40, 0.33]$& $[-1.127,0.314]$ & $[-0.11,-0.004]$ & $[-0.946,0.282]$ & $[0.047,0.107]$\tabularnewline\hline  
			$C_{10}^{'\text{NP}}$ & $[-0.25, 0.13]$ & $[-0.161,0.045]$ & $[0.004,0.11]$ & $[-0.218,0.065]$ & $[0.028,0.064]$\tabularnewline\hline  
			$C_{7}^{\text{NP}}$ &$[-0.01, 0.02]$& $[-7\times 10^{-5}, 0.024]$ & $[5.7\times 10^{-4},0.021]$ & $[-1.1\times 10^{-4},0.023]$ &$[5\times 10^{-5},8\times 10^{-5}]$\tabularnewline\hline   
			$C_{7}^{'\text{NP}}$    & $[0, 0.03]$  & $0 $ &0 & 0&0  \tabularnewline \hline 
			
			\hline 
		\end{tabular}
		\par
	\end{centering}

\end{table}
\section{\label{m5} Conclusion}
The extension of hypercharge introduces a family-nonuniversal extension of the SM, altering its phenomenology. Due to the non-universality of quark generations, the model requires additional Higgs doublets to generate quark masses and recover the CKM matrix. This non-universality  leads to FCNCs associated with both new vector gauge bosons and new scalar Higgs boson. 

The additional Higgs doublets involved in spontaneous symmetry breaking induce mixing between $Z$ and $Z^\prime$. This mixing can reduce the $Z_1$ mass compared to the SM  $Z$ mass and contribute positively to the $\rho$ parameter and $W$-boson mass.  Recent measurements of the $W$ boson mass  have shown a slight deviation from the SM  predictions, prompting further investigation.
Previous constraints on the $\rho$ parameter and $W$ mass measurement by CDF from Run II at the Tevatron  differed significantly other measurements,  ruling out the model with $x=\pm \frac{1}{6}$. However, the other measurements of $m_W$ \cite{ ATLAS:2024erm,LHCb:2021bjt} including the latest CMS result \cite{CMSMW2024} are in good agreement with each other and the SM prediction. By combining theses updated constraints on the $\rho$ parameter and $W$ mass, we find that the model can be viable for all cases of the $x$ parameter $x=\pm \frac{1}{2}, x=\pm \frac{1}{6}$. The analysis provides constraints on the new physics scale.  For instance, with positive $x=\left\{ 1/2,1/6 \right\}$, we respectively obtain $\La\in [4.3-28.1]$ TeV and $\La\in[13.2-39.3]$ TeV with $v_1\in [0,246]$ GeV. The cases with negative values $x=\left\{-1/2,-1/6\right\}$ are more complicated, with constraints like:
\begin{itemize}
\item For $x=-1/2$, we obtain the allowed regions $\La\in[1-16.7]$ TeV if $v_1\in [0, 204]$ GeV, and $\La \in [1-5.4]$ TeV if $v_1\in[ 220,246]$ GeV.
\item For $x=-1/2$,  the allowed regions are  $\La\in[1-5.7]$ TeV if $v_1\in[0,108]$ GeV , and $\La \in [1.5-16.7]$ TeV if $v_1\in [141,246]$ GeV.
\end{itemize}
We investigate both (axial)vector and (pseudo)scalar currents, including neutral and charged currents. In  lepton sector, the lepton flavor conserves at the tree-level. The charged scalar currents  contribute to leptonic observables such as BR$( e_i \to e_j \gamma)$, which are found to be highly suppressed due to the tight constraints on Yukawa coupling $h^{\nu}_{ij}\sim 10^{-5}$.

In quark sector, the FCNCs exit at the tree-level.
We consider the effects of both scalar and vector FCNCs on meson mixing systems $\Delta m_{K,B_s,B_d}$, branching ratios of top quark decays $t\to u(c)\gamma$, the $b\to s l^+ l^-$ clean observables BR$(B_s \to \mu^+ \mu^-)$, BR$(\bar{B}\to X_s\gamma$, and the $b\to sl^+l^-$ observables, which are strongly influenced by short-distance effects. 

We explore the parameter space that allows for explaining all mentioned flavor phenomenologies while remaining consistent with either  CDF measurement  \cite{CDF:2022hxs} or newest CMS measurement \cite{CMSMW2024} of the $W$ boson mass. Interestingly, we find that in both cases, the model with the charge parameter $x=1/2$ and NR mixing angles in $V_{d_R}$ matrix can not only accommodate the  constraints in Eqs. ( \ref{constraint1},\ref{constraint2}, \ref{Bsmm_constraint}, \ref{Bsmm_constraint}) but also other $b\to sl^+l^-$ observables. The other cases of $x$ are ruled out as they predict NP WCs outside the $1\sigma$ global fit interval  \cite{Alguero:2023jeh}.  Specifically, for the CDF case, we obtain the constraints $s_{12}^R\in[0.074,0.123]$, $\delta^R\in[0.52\pi,1.74\pi]$, $v_1\geq 10.6 $ GeV and $\La\geq 5.964$ TeV. For CMS case , the  constraints are broader:  $s_{12}^R\in[0,0.4]$, $\delta^R\in[0, 2\pi]$, $v_1\in [0,246] $ GeV and $\La\in [4.3-28.1]$ TeV.    
\section*{\label{acknowledgement}Acknowledgement}
This research is funded by Vietnam National Foundation for Science and Technology Development (NAFOSTED) under grant number 103.01-2023.50.

\bibliographystyle{apsrev4-1}
\bibliography{Refs}

\end{document}